\documentclass[sigplan,screen,authorversion]{acmart}

\copyrightyear{2026}
\acmYear{2026}
\setcopyright{cc}
\setcctype{by}
\acmConference[ASPLOS '26]{Proceedings of the 31st ACM International Conference on Architectural Support for Programming Languages and Operating Systems, Volume 2}{March 22--26, 2026}{Pittsburgh, PA, USA}
\acmBooktitle{Proceedings of the 31st ACM International Conference on Architectural Support for Programming Languages and Operating Systems, Volume 2 (ASPLOS '26), March 22--26, 2026, Pittsburgh, PA, USA}
\acmDOI{10.1145/3779212.3790215}
\acmISBN{979-8-4007-2359-9/2026/03}

\usepackage{listings}

\definecolor{javared}{rgb}{0.6,0,0} %
\definecolor{javagreen}{rgb}{0.25,0.5,0.35} %
\definecolor{javapurple}{rgb}{0.5,0,0.35} %
\definecolor{javadocblue}{rgb}{0.25,0.35,0.75} %

\AtBeginDocument{%
  \providecommand\BibTeX{{%
    \normalfont B\kern-0.5em{\scshape i\kern-0.25em b}\kern-0.8em\TeX}}}

\usepackage[]{hyperref}
\usepackage{multirow}
\usepackage{listings}
\usepackage{color}
\usepackage{svg}
\usepackage{amsmath}
\usepackage[normalem]{ulem}
\usepackage{fontawesome}
\usepackage{csquotes}
\usepackage{algorithm}
\usepackage{algpseudocode} 
\usepackage{wasysym} 
\usepackage{pdfcomment}

\begin{document}
\newcommand{\SQLiteSQLancer}{8.6K}

\newcommand{\ProjectName}{\emph{SQLancer++}}
\newcommand{\ProjectRepo}{\url{https://doi.org/10.5281/zenodo.18289297}} %
\newcommand{\DBMSUnderTestNum}{18}
\newcommand{\SQLancer}{\emph{SQLancer}}
\newcommand{\SQLancerSupportDBMSs}{22}
\newcommand{\SQLancerAvgLOC}{3,729}
\newcommand{\MethodFullName}{Adaptive Query Generator}
\newcommand{\MethodShorten}{AQG}
\newcommand{\SupportedDBMSs}{21}
\newcommand{\AvgLOCDBMS}{16}

\newcommand{\CPU}{64-core AMD EPYC 7763 CPU at 2.45GHz}
\newcommand{\ProjectLOC}{8.4K}
\newcommand{\SQLancerLOC}{83K}

\newcommand{\SQLiteValidRateWithFB}{97.7\%}
\newcommand{\SQLiteValidRateWOFB}{24.9\%}
\newcommand{\SQLiteValidRateSQLancer}{98.0\%}

\newcommand{\SQLiteValidRateIncrease}{292.5\%}
\newcommand{\PostgreSQLValideRateIncrease}{121.9\%}
\newcommand{\PostgreSQLValidRateWithFB}{52.4\%}
\newcommand{\PostgreSQLValidRateWOFB}{21.6\%}
\newcommand{\PostgreSQLValidRateSQLancer}{25.1\%}

\newcommand{\PCValidityRate}{XX\%}

\newcommand{\LogicBugs}{140}
\newcommand{\OtherBugs}{56}
\newcommand{\PerformanceBugs}{2}
\newcommand{\ErrorBugs}{20}
\newcommand{\CrashBugs}{34}
\newcommand{\TLPBugs}{133}
\newcommand{\NoRECBugs}{7}

\newcommand{\OverallBugs}{196}
\newcommand{\FixedBugs}{180}
\newcommand{\ConfirmedBugs}{12}
\newcommand{\DuplicatedBugs}{4}
\newcommand{\AvgCrateDBBugswithFB}{35.8}
\newcommand{\AvgCrateDBBugswoFB}{28.4}
\newcommand{\AvgCrateDBUniqueBugswithFB}{11.4}
\newcommand{\AvgCrateDBUniqueBugswoFB}{9.8}
\newcommand{\AvgPCAllBugs}{439.6}
\newcommand{\AvgPCUniqueBugs}{xx}

\newcommand{\WithAndWOFeedback}{$3.43\times$}
\newcommand{\WithFBSQLancer}{$3.01\times$}

\newcommand{\PostgresCrateDBInsertion}{309}
\newcommand{\PostgresCrateDBDeletion}{987}
\newcommand{\PostgresCrateDBModification}{1296}
\newcommand{\PostgresCrateDBErrorMessages}{33}
\newcommand{\PCDialectInsertion}{137}
\newcommand{\PCDialectDeletion}{709}
\newcommand{\PostgresCrateDialectModification}{846}
\newcommand{\PCSchemaInsertion}{XX}
\newcommand{\PCSchemaDeletion}{XX}

\newcommand{\FunctionNodes}{58}
\newcommand{\OperatorNodes}{47}

\newcommand{\SuccessfulBugsForMostDBMS}{8\%}
\newcommand{\OverallBugValidRate}{48\%}
\newcommand{\RowsContainInvalidCases}{13}
\newcommand{\ColumnsContainInvalidCases}{1}
\newcommand{\CrateDBBugsValidRepro}{16\%}
\newcommand{\VirtuosoValidRepro}{4\%}

\title{Scaling Automated Database System Testing}

\author{Suyang Zhong}
\email{suyang@u.nus.edu}
\orcid{0009-0003-0341-7362}
\affiliation{%
  \institution{National University of Singapore}
  \city{Singapore}
  \country{Singapore}
}

\author{Manuel Rigger}
\email{rigger@nus.edu.sg}
\orcid{0000-0001-8303-2099}
\affiliation{
    \institution{National University of Singapore}
  \city{Singapore}
    \country{Singapore}   
}

\renewcommand{\shortauthors}{Suyang Zhong and Manuel Rigger}

\begin{abstract}

Recently, various automated testing approaches have been proposed that use specialized test oracles to find hundreds of logic bugs in mature, widely-used Database Management Systems (DBMSs).
These test oracles require database and query generators, which must account for the often significant differences between the SQL dialects of these systems.
Since it can take weeks to implement such generators, many DBMS developers are unlikely to invest the time to adopt such automated testing approaches.
In short, existing approaches fail to scale to the plethora of DBMSs.
In this work, we present both a vision and a platform, \ProjectName{}, to apply test oracles to any SQL-based DBMS that supports a subset of common SQL features.
Our technical core contribution is a novel architecture for an \emph{adaptive SQL statement generator}.
This adaptive SQL generator generates SQL statements with various features, some of which might not be supported by the given DBMS, and then learns through interaction with the DBMS, which of these are understood by the DBMS.
Thus, over time, the generator will generate mostly valid SQL statements.
We evaluated \ProjectName{} across \DBMSUnderTestNum{} DBMSs and discovered a total of \OverallBugs{} unique, previously unknown bugs, of which \FixedBugs{} were fixed after we reported them.
While \ProjectName{} is the first major step towards scaling automated DBMS testing, various follow-up challenges remain.
\end{abstract}



\settopmatter{printfolios=false}
\maketitle

\section{Introduction}
Database Management Systems (DBMSs) are large, complex software systems. %
For example, MySQL consists of more than 5.5 million lines of code (LOC), and PostgreSQL has more than 1.7 million LOC.
Unsurprisingly, such DBMSs can be affected by bugs.
Automated testing approaches for DBMSs have been proposed to find so-called \emph{logic bugs}~\cite{slutz1998massive, rigger2020finding,rigger2020detecting,rigger2020testing,  ba2023testing, hao2023pinolo, liang2022detecting, tang2023detecting}, which are bugs that cause a system to silently compute an incorrect result, making such bugs notoriously difficult to find.
Many existing works aiming to find logic bugs proposed so-called \emph{test oracles} that can validate whether a query computes the correct result by transforming it in a semantic-preserving way and checking both queries' results' equivalence.
Overall, these approaches have found hundreds of bugs in widely-known DBMSs such as SQLite, MySQL, and PostgreSQL.

It would be ideal to apply automated DBMS testing approaches to the thousands of existing DBMSs.\footnote{See \url{https://dbdb.io/}}
The market for DBMSs is significant, currently being 162.25 billion USD and growing at a compound annual growth rate of 15.2\%~\cite{DatabaseSoftwareMarket2024}, fueling the development of new DBMSs, as well as further development of existing ones.
With the end of Moore's law,
various trends have set in posing new reliability challenges, such as the development of new, increasingly specialized DBMSs, often based on SQL and the relational model.
In addition, existing DBMSs are becoming increasingly complex, by using accelerators~\cite{korolija2022farview, huang2019xengine, sidler2017doppiodb}, %
or incorporating learned components~\cite{kraska2019sagedb, hilprecht2020learning, marcus2019neo}.
Overall, ensuring the correctness of DBMSs will become increasingly difficult.

\begin{figure}
    \centering
    \includegraphics[width=0.9\linewidth]{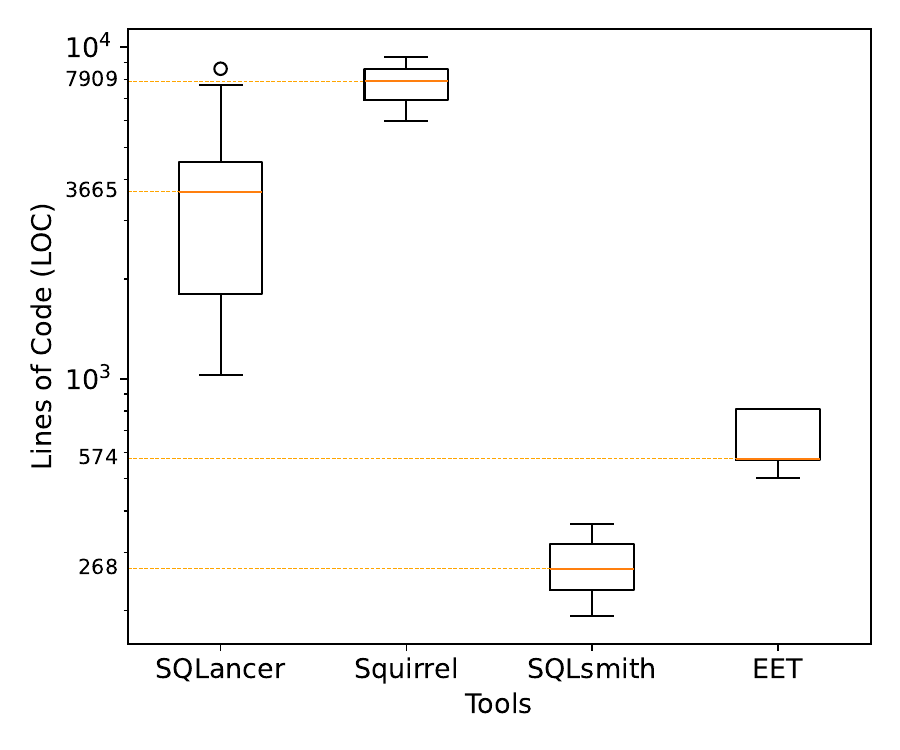}
    \caption{The lines of code needed in popular DBMS testing tools~\cite{rigger2020testing, zhong2020squirrel, sqlsmith, jiang2024detecting} for DBMS-specific components, where the orange line shows the median. Note that we only considered the core part of the generator of each DBMS.}
    \label{fig:LOC}
\end{figure}

\sloppy{}
Despite the significant, increasing need to apply automated testing approaches for DBMSs at scale, much effort is needed to implement and operate them.
A key challenge for current automated testing approaches is that test-case generators, which generate SQL statements that create a database, populate it with data, and execute queries, must be manually implemented to account for the differences in SQL dialects~\cite{rigger2020detecting,sqlsmith}.
Doing so typically requires implementing thousands of lines of code. %
For example, SQLancer~\cite{rigger2020detecting, rigger2020finding, rigger2020testing, ba2023finding, ba2023testing}, a state-of-the-art tool for DBMS testing, currently supports generators for \SQLancerSupportDBMSs{} DBMSs, which, on average, are implemented in \SQLancerAvgLOC{} LOC (see Figure~\ref{fig:LOC}), with some DBMS-specific components being contributed by major companies.
However, most DBMS development teams are unlikely to invest this effort.
For example, a Vitess blog post describes that they considered using SQLancer, but realized that \emph{``It would take a lot of work to properly integrate Vitess with SQLancer, due to each DBMS tester in SQLancer essentially being written completely separately with similar logic.''}~\cite{vitess2024fuzzing}
While Vitess aims to be MySQL-compatible, and SQLancer provides a MySQL implementation, in practice, various differences still exist that make it difficult to reuse generators in such a scenario.

In this work, we introduce our vision of applying testing DBMSs \emph{at scale}.
We propose a new automated testing platform, \ProjectName{}, as part of our ongoing effort to reduce the effort of implementing and operating DBMS testing approaches.
The main technical contribution of this work is an adaptive query generator, which infers to generate statements that are understood by the DBMS under test.
This generator repeatedly sends SQL test cases to the DBMS; if a SQL feature is not understood, the generator will prevent its generation and thus, over time, increase the validity rate of generated statements.
Such a generator is unlikely to perform as well as one that is implemented for a specific system; however, it can be directly applied without additional implementation effort.
In addition, any added features might also be immediately applicable to other DBMSs.
Other key design decisions include an internal model of the database schema, to eschew avoiding querying schema information from DBMSs, which often requires using DBMS-specific interfaces.
Finally, we propose a synergistic and pragmatic bug-prioritization approach, which reports bug-inducing tests only if its features are not present in previously reported bugs.
Overall, we believe that these techniques are both simple and practical.

Our results are promising, as we have found and reported \OverallBugs{} unique bugs across \DBMSUnderTestNum{} DBMSs, demonstrating that the approach is effective and scalable.
As our bug-finding efforts on these systems have not yet saturated, we are still reporting more bugs.
\FixedBugs{} bugs have already been fixed, indicating that the developers considered the bugs important.
The feedback mechanism in the adaptive generator significantly increased the validity rate of statements; for SQLite, by \SQLiteValidRateIncrease{}. 
Similarly, the prioritization approach has shown promise.
In CrateDB, \ProjectName{} identified over 60K bug-inducing cases in an hour on average, which the bug prioritizer reduced to \AvgCrateDBBugswithFB, with \AvgCrateDBUniqueBugswithFB{} being unique bugs.

\ProjectName{} is the first step towards an automated testing platform for DBMSs that can find bugs at a large scale, similar to platforms such as OSS-Fuzz~\cite{kostya2017oss} and syzkaller~\cite{syzkaller}, which tackled the problem of scaling general-purpose greybox fuzzers like AFL, as well as kernel fuzzing.
Overall, we hope that \ProjectName{} will have a significant impact in improving the reliability of DBMSs at large and, in particular, help smaller DBMS development teams that cannot invest significant resources into testing.
To facilitate future research and adoption, we have released \ProjectName{} at \ProjectRepo{}.

In summary, we propose the following:
\begin{itemize}
    \item At a conceptual level, we have identified the problem of scaling existing DBMS testing approaches and propose a new testing platform, \ProjectName{} to apply previously-proposed test oracles at scale.
    \item At a technical level, we propose an adaptive query generator, a model-based schema representation, as well as a bug-prioritization approach for logic bugs.
    \item At an empirical level, we show that these techniques are highly effective, and already found \OverallBugs{} unique, previously unknown bugs in popular DBMSs.
\end{itemize}

\section{Background and Motivation}
\label{sec:sqlancer-intro}
\SQLancer{} is among the most popular automated testing tools for DBMSs, supports over 20 DBMSs, and is widely used by DBMS developers.
While the \SQLancer{} authors proposed various approaches built on it~\cite{rigger2020detecting,rigger2020finding,rigger2020testing, ba2023finding, ba2023testing,ba2024keep}, also other tools have been prototyped on it~\cite{song2023testing,PingCAP-Qe, song2024detecting, liu2024conformance}.
Thus, we focus on \SQLancer{} to motivate the limitations of existing approaches. %
Different from other fuzzing approaches, such as SQLsmith, \SQLancer{} implements various state-of-the-art test oracles~\cite{rigger2020detecting, rigger2020finding, rigger2020testing, ba2023finding, ba2023testing} for finding logic and performance issues.

\paragraph{SQL generators}
At the core of automated DBMS testing tools like \SQLancer{} are DBMS-specific, rule-based SQL generators. Listing~\ref{listing:duckdb-example} shows a simplified excerpt of a \SQLancer{} generator for \texttt{CREATE INDEX} statements. 
Line 2 uses a string to represent a \texttt{CREATE} statement, to which, subsequently, other elements such as \texttt{INDEX} are added.
To create an index, schema information is required. Specifically, line 7 determines an index name that is not yet present in the database, and line 9 determines an existing table, on which the index should be created.
\SQLancer{}---as well as other DBMS testing tools like SQLsmith or Griffin---query this information from the DBMS under test.
As shown by the call to \texttt{random.nextBoolean()} in lines 3 and 11, it is randomly decided whether to create a unique index by including the \texttt{UNIQUE} keyword, or a partial index by including the \texttt{WHERE} keyword; typically, in \SQLancer{}, the probabilities of choosing a feature are uniformly distributed.

\begin{figure}[tb]
    \captionof{lstlisting}{Simplified excerpt of \texttt{PostgresIndexGenerator}. }
    \label{listing:duckdb-example}
\begin{minipage}{.45\textwidth}
    \centering
    \lstset{language=Java,xleftmargin=2em, lineskip=0pt, aboveskip=0pt,belowskip=0pt,basicstyle=\ttfamily\scriptsize }
\begin{lstlisting}
public SQLStatement createIndex(PostgresState state) {
  String stmt = "CREATE ";
  if (state.nextBoolean()) {
    stmt += "UNIQUE ";
  }
  stmt += "INDEX ";
  stmt += state.getSchema().getFreeIndexName();
  stmt += " ON ";
  stmt += state.getSchema().getRandomTableName();
  stmt += "(" + table.getRandColumnNames() + ")";
  if (state.nextBoolean()) {
    stmt += " WHERE ";
    stmt += state.getRandomExpr(table.getColumns());
  }
  return new SQLStatement(stmt);
}
\end{lstlisting}
\end{minipage}
\end{figure}

\paragraph{Case study}
To motivate the challenges, we assumed that the CrateDB development team would want to adopt one DBMS testing tool to test their system.
CrateDB is a PostgreSQL-compatible system; luckily, a PostgreSQL generator already exists within \SQLancer{}, and it would be reasonable to assume that the CrateDB developers would want to adapt it, rather than creating one without reference.
However, even given this best-possible scenario, our case study demonstrates that the adaptation effort is still high.
In order to minimally adapt the generator, we added \PostgresCrateDBInsertion{} and deleted \PostgresCrateDBDeletion{} LOC, resulting in \PostgresCrateDBModification{} LOC being modified to avoid syntax errors and acquire meta-data for schema information.
Performing these changes is non-trivial, requiring detailed knowledge of \SQLancer{}.
Several other DBMS testing tools exist, such as \emph{Griffin}~\cite{fu2023griffin} for grammar-free testing, and \emph{SQLRight}~\cite{liang2022detecting} for detecting logic bugs. However, both tools were built upon AFL~\cite{AFL}, making them inapplicable to DBMS written in Java, like CrateDB.
While Java-based fuzzers like JQF~\cite{padhye2019fuzzing} or Jazzer~\cite{jazzer} could be used for Java-based DBMSs, these fuzzers are agnostic to SQL, and would thus struggle with generating meaningful sequences of SQL statements.

\paragraph{C1. SQL dialects}
First, the SQL generators are highly DBMS-specific as they must account for differences in statements, operators, functions, and data types.
The above generator is specific to PostgreSQL, which supports partial indexes, which are, for example, unavailable in MySQL.
The full versions of such generators typically contain many uncommon keywords, testing which is desirable, as unique functionalities are prevalent and might be affected by bugs.
For example, index generators might generate additional DBMS-specific index types (\emph{e.g.}, \texttt{HASH} indexes in PostgreSQL), or use DBMS-specific operators in expressions. %
Thus, applying \SQLancer{} or other DBMS testing tools to a new DBMS requires significant implementation effort in implementing new generators, as shown in Figure~\ref{fig:LOC}.
Regarding our case study, despite CrateDB claiming to be dialect-compatible with PostgreSQL, we had to modify \PostgresCrateDialectModification{} LOC to avoid syntax errors.
If these syntax errors were simply ignored, an experiment we performed suggests that the validity rate of queries would drop to less than 1\%, significantly affecting \SQLancer{}'s performance.
Going beyond a minimal implementation, the CrateDB developers would subsequently likely also need to add CrateDB-specific functionality, incurring additional effort.

\paragraph{C2. Schema state}
Second, acquiring metadata for schema information differs across DBMSs.
In SQLite, \texttt{sqlite\_master} holds the schema information. 
However, PostgreSQL, for example, exposes similar information in \texttt{information\_schema.tables}, and MySQL-like systems provide SQL statements like \texttt{SHOW TABLE}.
In our case study, we found that even when DBMSs share the same metadata tables (\emph{e.g.}, \texttt{information\_schema} in CrateDB), the attributes of these tables can differ, requiring the logic that parses the schema to be changed.
Besides, several features are related to metadata tables that are not fully supported by CrateDB (\emph{e.g.}, collation information in \texttt{pg\_collation}), and we manually omitted these.
Note that, as with \emph{SQLancer}, several other tools---such as Griffin~\cite{fu2023griffin} and DynSQL~\cite{jiang2023dynsql}---also acquire schema state through a similar interface.

\paragraph{C3. Duplicate bug-inducing test cases}
Third, \SQLancer{} has no deduplication mechanism and is likely to repeatedly trigger the same underlying bugs.
This is problematic since manual analysis is necessary to determine whether two bug-inducing test cases likely trigger the same underlying bug, to avoid overburdening developers by reporting duplicate bugs.
In our case study, running the generator on a historic version of CrateDB for one hour would result in more than 400 bug-inducing test cases.
Whether two bug-inducing test cases trigger the same logic bug is decided by the developers' verdict; it is not an inherent property of the test cases, meaning that the problem cannot be fully addressed.
Existing techniques such as deduplication based on stack traces cannot be directly applied to logic bugs~\cite{chen2013taming, jiang2021igor}.
Despite this, we believe a pragmatic solution is required to deal with this challenge in practice.

\section{SQLancer++}
\begin{figure}[tb]
    \centering
    \includegraphics[width=\linewidth]{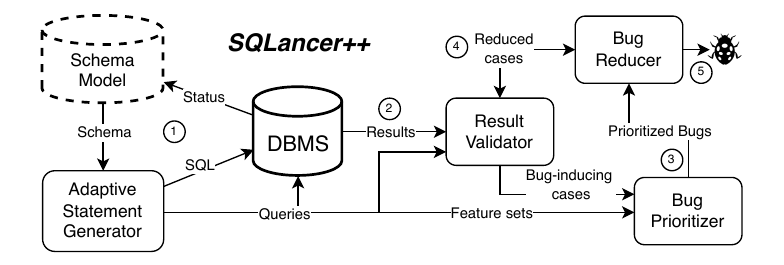}
    \caption{Architecture of the \ProjectName{}}
    
    \label{fig:Project-overview}
\end{figure}

\ProjectName{} is an automated DBMS testing platform aiming to find logic bugs in DBMSs (see Figure~\ref{fig:Project-overview}), addressing the scalability limitations of existing rule-based approaches.
The core component of \ProjectName{} is an \emph{adaptive statement generator}, which infers the supported SQL features of the target DBMS during execution and adaptively generates random SQL statements that can be processed by the DBMS under test.
First, the adaptive generator generates SQL statements to create a database state while maintaining an internal model of the schema.
Then, the generator generates random queries and executes them on the DBMS, and a feedback mechanism automatically prioritizes the supported features and deprioritize the unsupported features to ensure most queries are semantically valid.
After retrieving the queries' results, an oracle validator checks whether the results are as expected.
When detecting issues, \ProjectName{} compares the SQL features with the previously found bug-inducing test cases to prioritize the bug-inducing test cases for reporting.
If a previous bug-inducing feature set is a subset of the new test case, it is marked as a potential duplicate.

\paragraph{SQL features}
The notion of SQL \emph{features} is an important, but abstract concept in both the generator and bug prioritizer.
A feature refers to an element or property in the query language, which we expect to be either supported or unsupported by a given DBMS.
More concretely, developers can mark any code block in a generator in \ProjectName{} as a potential feature. This instructs the generator to determine whether the feature is supported by the DBMS (\emph{i.e.}, whether statements based on which the feature is generated execute successfully), and only if so, continue generating the feature. In addition, it instructs the prioritizer to use it for bug-prioritization.
Features can be specified in different granularities.
Often, a feature might be a specific keyword or operator.
However, it can also refer to a class of functions, for example, string functions.
Finally, it can be used to specify abstract properties, such as whether the DBMS provides implicit conversions between most types.

\paragraph{Adaptive statement generator} 
The core of \ProjectName{} is an adaptive statement generator detailed in Section~\ref{sec:aqg}, which tackles C1, the differences across DBMSs' SQL dialects.
The generator produces SQL statements with SQL features that might not be supported by all DBMSs, or, whose constraints are difficult to meet, causing statements containing them to frequently fail. 
It learns through execution feedback from the DBMS under test.
Users can set a minimum success probability threshold for a feature, which is estimated through \emph{Bayesian inference}~\cite{gelman1995bayesian} by utilizing feedback from previous executions. This suppresses the generation of frequently failing SQL features to increase the validity rate of SQL statements.

\begin{figure}[tb]
    \centering
    \includegraphics[width=\linewidth]{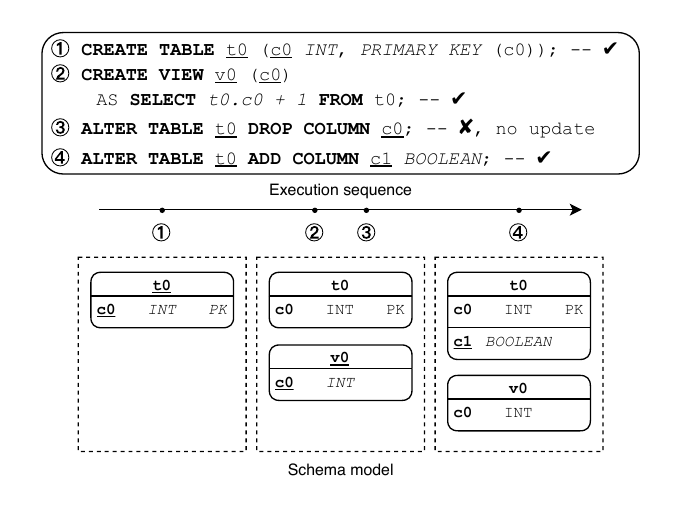}
    \caption{Schema model after executing each DDL statement}
    \label{fig:ddl-schema}
\end{figure}

\paragraph{Schema model}
To address C2, the challenge of obtaining database schemas, the generator maintains an internal data model for the database, enabling it to obtain schema information without querying the DBMS.
The internal data model matches the DBMSs' model by simulating the behavior of DDL statements generated and checking the execution status from the DBMS.
Initially, the objects (\emph{i.e.}, tables, views, or indexes) in the model are $\mathcal{O} = \{\}$.
When executing a statement $s$ that potentially creates an object $T$, such as a \texttt{CREATE TABLE} statement creating a table with a specific name, we obtain the execution status from the DBMS.
If the object was successfully created, we add the object to our model of the schema such that  $\mathcal{O} = \mathcal{O} \cup T$; otherwise, we consider the creation of the table $T$ unsuccessful.

As shown in Figure~\ref{fig:ddl-schema}, when generating a \texttt{CREATE TABLE} statement (\textcircled{1}), we create an internal schema model of the table, which records its name (\texttt{t0}) along with its columns (\texttt{c0}) and corresponding data types (\texttt{INT}). 
Once the statement is successfully executed on the DBMS, this abstract object is added to the schema model.
During subsequent testing, the schema model can be queried to retrieve available tables and their columns.
Potential bugs could cause DBMSs to maintain an incorrect schema state, in which case our internal schema model could deviate from it.
However, during our fuzzing campaign, we have not observed any related false positives.

\paragraph{Result validator}
\ProjectName{} can be combined with any test oracle that is not specific to a DBMS. 
We adopted two state-of-the-art~\cite{gao2023comprehensive} test oracles, \emph{Ternary Logic Partitioning} (TLP)~\cite{rigger2020finding} and \emph{Non-optimizing Reference Engine Construction} (NoREC)~\cite{rigger2020detecting}, which can find logic bugs by comparing the results of two equivalent queries constructed through a syntactic transformation that applies to any DBMS. 
The validator fetches the results of two equivalent queries and validates if the results are the same. If not, it reports a bug.

\paragraph{Bugs prioritizer}

\begin{figure}[tb]
    \centering
    \includegraphics[width=\linewidth]{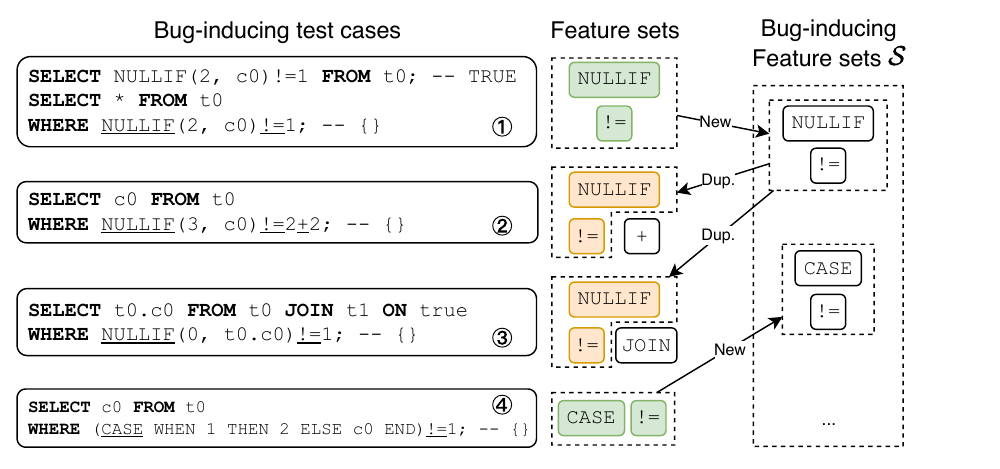}
    \caption{Bug prioritization. We omitted the unoptimized queries in test cases \textcircled{2} to \textcircled{4} for simplicity.}
    \label{fig:deduplication}
\end{figure}

\ProjectName{} employs a bug prioritization approach to prioritize bug-inducing test cases for analysis and reporting, aiming to minimize the amount of duplicate bugs being reported, tackling C3.
We prioritize the bugs by comparing SQL feature sets from previous bug-inducing test cases with the feature set from the newly found bug-inducing test case.
We define the \emph{feature set} $S$ as a set of features $\{F_1, F_2, \cdots, F_n\}$, where each $F_i$  denotes one SQL feature that was enabled when generating a SQL statement. 
Specifically, assume that $\mathcal{S}$ represents the bug-inducing feature sets of the historical bugs, and $S_{new}$ denotes the feature set of the newly detected bug-inducing test case. Should there exist an $S$ within $\mathcal{S}$ such that $S \subseteq S_{new}$, then this test case is considered a potential duplicate, based on the intuition that the root cause of the bugs might be the faulty implementation of the SQL features in the bug-inducing test cases triggering the bugs.
Figure~\ref{fig:deduplication} shows an example of the bug prioritization.
When triggering a bug with feature set $S_{new}=\texttt{\{NULLIF, !=\}}$, we retrieve $\mathcal{S}$ and find no set is a subset of $S_{new}$. Thus, we consider this a new bug and update $\mathcal{S}$.
Subsequently, feature sets of test cases \textcircled{2} and \textcircled{3} contain \texttt{\{NULL, !=\}}, and we mark these cases as potential duplicated, meaning that would analyze and report them only once existing bugs have been fixed by the developers.

Given two bug-inducing test cases, the generator might misclassify them both as triggering the same underlying bug when they are not, or misclassify them as triggering different bugs when they trigger the same.
For example, the expression \texttt{NULLIF(2, c0)<>1} may trigger a bug involving the feature set \texttt{\{NULLIF, <>\}}, which would be considered new although its root cause might be identical to a previously reported issue.
However, we believe our method is effective for reducing the large volume of reported bugs---potentially hundreds or thousands in one hour---when applying \ProjectName{} to an untested system (see Section~\ref{sec: bug-deduplication}).
False negatives can occur, which means two different bugs affect the statements with the same feature sets.
However, these do not result in missed bugs; after developers fix the prioritized bugs, they can run the reproducer to reproduce the detected bugs again quickly to check if there are any false negatives.

\section{Adaptive Statement Generator}\label{sec:aqg}\label{sec:approach}

\begin{figure*}[tb]
    \centering
    \includegraphics[width=\linewidth]{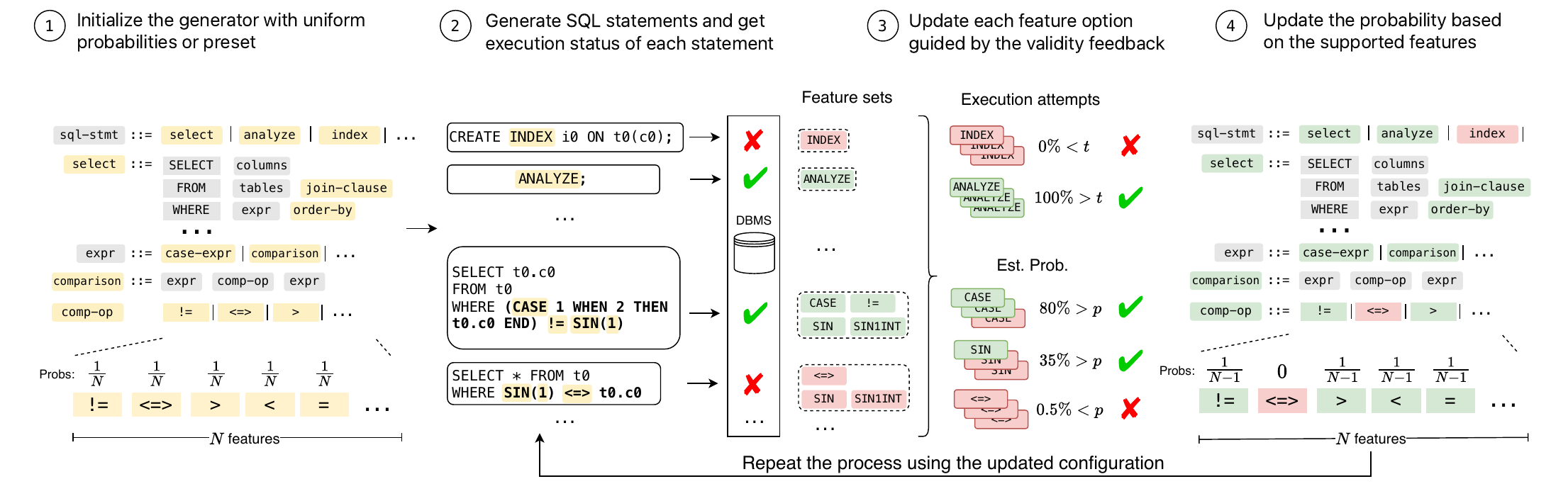}
    \caption{Overview of the adaptive statement generator}
    \label{fig:feedback}
\end{figure*}

Figure~\ref{fig:feedback} illustrates the steps of adaptive statement generation, which is at the core of \ProjectName{}.
First, we initialize the generator with each feature sharing the same probability as the other options in the respective context (see step \textcircled{1}).
Next, we randomly generate statements and queries, whose results are checked by the provided test oracle.
Expectedly, some of these statements might contain features that are not supported by the DBMS under test.
To identify these, we generate each statement one by one, execute it on the DBMS, and obtain its execution status, which indicates whether the statement could be executed successfully, whether it failed (see step \textcircled{2}).
Each feature in the feature set is marked as valid (shown in green) if the statement succeeds, and as failed otherwise (shown in red).
We repeat this process for an empirically-determined number of iterations.
In step \textcircled{3}, we calculate the estimated probability of successfully executing each SQL feature, using \emph{Bayesian inference}~\cite{gelman1995bayesian} based on previous execution feedback.
Features are marked as \emph{unsupported}---unlikely to execute successfully---for the target system if their estimated probability falls below a specified threshold; otherwise, they will be regarded as \emph{supported}.
In step \textcircled{4}, we update the generator based on the state of each feature.
Unsupported features are subsequently suppressed, and the probabilities for other alternatives are distributed uniformly. %
The probabilities from step \textcircled{4} can be persisted in a file and loaded in step \textcircled{1} of future executions.

\paragraph{SQL statement generation}
The generator generates random SQL statements based on the specified rule and records the corresponding feature set. %
By default, the selection of each feature under each grammar rule is uniformly distributed (see step \textcircled{1}).
For example, \texttt{<=>} is a feature in the generator for generating a null-safe comparison expression in the SQL statement. 
Correspondingly, we add this feature to the feature set such that $S=S\cup \{\texttt{<=>}\}$.

\paragraph{Validity feedback}
We execute the generated statements in step \textcircled{2} and record their execution statuses as feedback. 
If the DBMS encounters a syntax or semantic error---whether caused by unsupported features, constraints, or any other issue---it returns an error message, causing that statement to fail.
For example, in Figure~\ref{fig:feedback}, the DBMS under test does not support indexes, which is why the first statement fails. 
The third SQL statement executes successfully, and it contains features \texttt{CASE}, \texttt{!=}, \texttt{SIN} and \texttt{SIN1INT}. 
Here, \texttt{SIN1INT} is a composite feature for data types, which we will describe later.
Consequently, we mark the feature set \texttt{\{INDEX\}} as \emph{failed} and \texttt{\{CASE, !=, SIN, SIN1INT\}} as \emph{succeeded}.
The same feature might fail in one context but succeed in another (\emph{e.g.}, \texttt{ASIN(1)} can succeed in PostgreSQL while \texttt{ASIN(2)} will throw an error). 
If a feature that rarely results in a valid SQL statement is generated frequently, it lowers the overall success rate of statements, preventing the DBMS's core logic from being exercised, which is critical for finding logic bugs.
Thus, the next steps aim to analyze the feature sets to infer whether features are \emph{supported} (\emph{i.e.}, executed successfully).

\paragraph{Feedback mechanism}
In step \textcircled{3}, we estimate the probability that the respective features are supported by the DBMS via validity feedback.
Since our bug detection operates in two phases---first establishing the database state via DDL (Data Definition Language) or DML (Data Manipulation Language) statements, and then issuing a large number of random queries---we evaluate features in these two categories separately.
For \texttt{DDL/DML} statements, we apply straightforward tests: if a feature repeatedly fails beyond a user-specified number of attempts, it is deemed unsupported. 
For queries, we employ a simple \emph{Bayesian inference}~\cite{gelman1995bayesian} model to estimate the posterior probability distribution indicating whether each query feature is \emph{supported}. 
Users may specify a threshold $p$ (\emph{e.g.}, $1\%$) for the least query successful probability, and if the posterior probability mass for a given feature predominantly falls below this threshold, \ProjectName{} marks this feature as unsupported.

\paragraph{Statistical model}
For each iteration, we record each feature's total number of executions $N$ and successful executions $y$.
Queries are generated independently, unlike test case generation methods with guidance (\emph{e.g.}, code coverage). 
Consequently, each feature has a constant probability of being included in any given query. 
Assuming that every feature behaves deterministically and leaves the database state unaltered, the probability of success for each feature is independent and identically distributed among each iteration.
This scenario can be modeled using a binomial distribution, where the probability mass function is defined as
\begin{equation}\label{binom}
    p(y|\theta) = \binom{N}{y} \theta^y(1-\theta)^{N-y}
\end{equation}
with $\theta$ representing the probability that the feature is executed successfully. 
To assess whether a feature can be executed successfully, we calculate the posterior probability distribution $p(\theta|y)$. 
In the worst-case scenario, we assume a uniform prior distribution for $\theta$ over the interval $[0,1]$, implying that $p(\theta)=1$.
Applying Bayes' theorem, we obtain
\begin{equation}\label{eq:bayes}
    p(\theta|y) = \frac{p(y|\theta)\cdot p(\theta)}{p(y)}
\end{equation}
where $p(y)=\int_0^1 p(\theta)p(y|\theta)d\theta$. 
Substituting Equation~\ref{binom} into Equation~\ref{eq:bayes}, we find
\begin{equation}
    p(\theta|y)\propto  \theta^y(1-\theta)^{N-y}
\end{equation}
indicating that $\theta|y$ follows a Beta distribution: $\theta|y\sim\text{Beta}(y+1,N-y+1)$.
Consider a user-specified threshold $p=0.01$, which indicates that the success probability should be at least $1\%$. 
In the case where $y=0, N=400$---meaning the feature has been executed 400 times without any successful executions---the posterior distribution is then $\text{Beta}(1,401)$. The $95\%$ credible interval of this posterior is approximately $[6\times10^{-5}, 0.009]$.
Since more than $95\%$ of the posterior probability mass is below $0.01$, the feature is deemed unsupported.
Lowering the threshold $p$ would necessitate additional executions to achieve a similar level of confidence.

\paragraph{Generator specification}
In step \textcircled{4}, we update the probability of the generator choices based on the states of the SQL features.
Specifically, our goal is to prevent unsupported features from being generated.
For example, the comparison operator \texttt{``<=>''} is not supported for the DBMS under test in the example shown in Figure~\ref{fig:feedback}.
We assign zero probability to \texttt{``<=>''}.
Considering that the total number of features in this production rule is $N$, and that the other alternatives are uniformly distributed, each receives a probability of $1/(N - 1)$.
Based on the updated rule in step \textcircled{4}, the generator has a higher probability of selecting other comparison operators, avoiding generating unsupported \texttt{``<=>''}, which increases the validity rate of statements.

\section{Evaluation}
We sought to understand how effective and efficient \ProjectName{} is in finding bugs in DBMSs at a large scale from these perspectives: the overall bug-finding effectiveness (see Section~\ref{sec: bug-detect}), the impact of SQL features (see Section~\ref{sec:featurestudy}), bug-finding efficiency (see Section~\ref{sec: bug-efficiency}), feedback mechanism effectiveness (see Section~\ref{sec: ablation-study}), and bug-prioritization effectiveness (see Section~\ref{sec: bug-deduplication}).

\begin{table}
    \centering
\footnotesize
\caption{Comparison of related DBMS testing approaches. Base tools are shown in bold. \emph{Eval.} denotes the number of DBMSs reported in the research paper. \Circle{} indicates close-to-zero manual effort for adapting the tool to a new DBMS, while the \CIRCLE{} indicates that manual adaptation is necessary.}
\label{tab:related-work}
\setlength{\tabcolsep}{4pt}

\begin{tabular}{lccccrc}
\toprule
\multirow{2}{*}{Tool}       & \multicolumn{2}{c}{Bug Type}&  \multicolumn{3}{c}{DBMS}        &  Manual  \\
                                & Crash & Logic                 & C/C++  & Non C    & Eval.      & Efforts   \\
\midrule
\emph{\textbf{AFL}}~\cite{AFL}        & \checkmark & - & \checkmark & -  & - & \Circle \\
- \emph{Griffin}~\cite{fu2023griffin}  & \checkmark & - & \checkmark & - & 4 & \Circle \\
- \emph{WingFuzz}~\cite{liang2024wingfuzz}  & \checkmark & - & \checkmark & - & 12 & \Circle \\
- \emph{SQLRight}~\cite{liang2022detecting} & \checkmark & \checkmark & \checkmark & -  & 3 & \CIRCLE \\
\emph{\textbf{SQLSmith}}~\cite{sqlsmith}  & \checkmark & - & \checkmark & \checkmark & - & \CIRCLE \\
- \emph{EET}~\cite{jiang2024detecting}   & \checkmark & \checkmark & \checkmark & \checkmark & 5 & \CIRCLE \\
\emph{\textbf{SQLancer}}~\cite{rigger2020testing} & \checkmark & \checkmark & \checkmark & \checkmark & - & \CIRCLE \\
- \emph{TLP}~\cite{rigger2020finding} & \checkmark & \checkmark & \checkmark & \checkmark & 6 & \CIRCLE \\
- \emph{NoREC}~\cite{rigger2020detecting} & \checkmark & \checkmark & \checkmark & \checkmark & 4 & \CIRCLE \\
- \emph{CODDTest}~\cite{zhang2025constant} & \checkmark & \checkmark & \checkmark & \checkmark & 5 & \CIRCLE \\
\ProjectName{} & \checkmark & \checkmark & \checkmark & \checkmark & 18 &\Circle \\
\bottomrule

\end{tabular}

\end{table}

\paragraph{Baselines}
We implemented \ProjectName{} in Java in \ProjectLOC{} LOC. 
We compared \ProjectName{} with \SQLancer{} and \emph{SQLRight}, which are state-of-the-art logic bug detection tools.
In comparison, \SQLancer{} has \SQLancerLOC{} lines of Java code, and the core of \emph{SQLRight} has 30K lines of C++ code.
Table~\ref{tab:related-work} demonstrates the other DBMS testing tools.
Our aim for \ProjectName{} was to detect logic bugs in various DBMSs with different dialects; thus, for a given DBMS that is supported by \SQLancer{}, we would expect it to perform better than \ProjectName{} due to its manually-written generators specific to that system. Our main use case is to apply \ProjectName{} to SQL dialects that are not supported by existing DBMSs.
We did not compare with other AFL-based fuzzers, including Griffin~\cite{fu2023griffin}, WingFuzz~\cite{liang2024wingfuzz}, and BUZZBEE~\cite{yang2024towards}, since they cannot detect logic bugs or require instrumentation, which is limited to C/C++ systems, whereas many DBMSs are written in other languages (\emph{e.g.}, CrateDB is written in Java).

\paragraph{Experiments Setup}
We conducted the experiments using a server with a \CPU{} and 512GB memory running Ubuntu 22.04. 
In the bug-finding campaign, we enabled multi-threading; however, for other experiments, we evaluated \ProjectName{} using a single thread to prevent feedback sharing across threads, as this could mask the specific impact of individual threads.
We set the maximum expression depth to three and randomly created up to two tables and one view, respectively, which are the standard settings for \SQLancer{}.
We also compared \ProjectName{} under different feedback mechanisms. ``\ProjectName{}'' denotes the enabling of feedback, and ``\emph{SQLancer++\textsubscript{Rand}}'' indicates the absence of feedback guiding the generator.

\paragraph{DBMS selection}
We considered \DBMSUnderTestNum{} DBMSs in our evaluation (see Table~\ref{table:DBMSs}).
We first included DBMSs that had been extensively studied in prior work (DuckDB, H2, MariaDB, MonetDB, MySQL, Percona MySQL, SQLite, TiDB, and Virtuoso). We then added systems that were not covered by \SQLancer{}, but were popular according to DB-Engines rank (at most 50) or GitHub stars (at least 4K), including Firebird, Oracle, CrateDB, Dolt, RisingWave, and Vitess, and further included Umbra and CedarDB as prominent recent academic systems.
To test the effectiveness of \ProjectName{} (see Section~\ref{sec: bug-detect}), we tested them in a bug-finding campaign, using their latest available development versions to avoid reporting bugs that had already been fixed.
For bug-prioritization effectiveness and bug-finding efficiency, we aimed to select DBMS that could be continuously tested for a longer period while also having a number of bugs that are fixed in the latest version, allowing us to determine whether a bug was unique based on its fix commit. CrateDB met these criteria, as all bugs we found are logic bugs, and we selected CrateDB 5.5.0, a historic version, for testing.
To compare with other tools, we selected SQLite 3.45.2 as the baseline for measuring coverage and validity rate and selected PostgreSQL 14.11 for measuring validity rate. To the best of our knowledge, no logic bug-finding tools could be applied to CrateDB. 
SQLite and PostgreSQL are robust systems, which have been extensively tested and targeted by many bug-finding tools~\cite{rigger2020detecting, zhong2020squirrel}.
We did not compare \ProjectName{} with \SQLancer{} by testing MySQL, although both tools could find bugs in it, since most bugs found by previous automated testing work remain unfixed~\cite{hao2023pinolo, rigger2020testing, gao2023comprehensive}.

\subsection{Effectiveness}\label{sec: bug-detect}
We continuously tested the \DBMSUnderTestNum{} DBMSs for about four months of intensive testing, followed by several months of intermittent testing.
Typically, we ran \ProjectName{} for several seconds up to multiple hours, and then further processed the automatically-reduced and prioritized bug-inducing test cases.
Before reporting them to the developers, we further reduced them manually and checked whether any similar bugs had already been reported to avoid duplicate issues.
For the developers of DBMSs who were actively fixing bugs, we reported up to three bug-inducing test cases.
For some DBMSs where bugs were not fixed, like for MySQL (as also pointed out by prior work~\cite{rigger2020detecting, rigger2020finding}), we refrained from reporting bugs.
After previous bugs were fixed, we started another testing run.
Similar testing campaigns have also been conducted to evaluate the effectiveness of testing approaches for DBMSs~\cite{rigger2020testing, jiang2023detecting, liang2022detecting} as well as in other contexts~\cite{lecoeur2023program, le2014compiler, zhang2017skeletal}.

\begin{table*}[tb]
    \centering
\small
\caption{\ProjectName{} allowed us to find and report \OverallBugs{} bugs in \DBMSUnderTestNum{} systems, of which \FixedBugs{} were fixed by the developers, and of which \LogicBugs{} were logic bugs. DBMSs are sorted alphabetically.}\label{table:DBMSs}
\begin{tabular}{ lrrrrrrrrrrcc }
\toprule
DBMS & \multirow{2}{*}{All}  & \multicolumn{3}{c}{Bug Status} & \multicolumn{2}{c}{Test Oracle} & DB-Engines & GitHub & Released/ & Lines of &SQLancer & C/C++ \\
Names &                        &       Fixed        &       Conf.       &      Dupl.        & Logic & Other&Rankings   & Stars  & Published & Code & Support & System \\
\midrule
CedarDB    &  4   &  4  & 0 & 0  & 1 & 3  &  -  & -     & 2024 &  -  & -   &      $\checkmark$                    \\
CrateDB    &  28  &  26 & 0 & 2 & 28 & 0  & 229 & 4.0k  & 2017 &  597K  & -  &    -                      \\
Cubrid     &  1   &  1  & 0 & 0  & 1  & 0  & 169 & 0.3k  & 2008 & 1105K & -   & $\checkmark$               \\
Dolt       &  28  &  27 & 1 & 0 & 16 & 12 & 197 & 16.9k & 2018 & 380K & -    &       -                         \\
DuckDB     &  10   &  9  & 1 & 0  & 6  & 4  & 76  & 16.4k & 2019 & 1496K & $\checkmark$  & $\checkmark$                \\
Firebird   &  11  &  9  & 1 & 1  & 9  & 2  & 31  & 1.2k  & 2000 &  1643K & -     &   $\checkmark$            \\
H2         &  2   &  2  & 0 & 0  & 1  & 1  & 50  & 4.0k  & 2005 & 297K & $\checkmark$   & -               \\
MariaDB    &  2   &  0  & 2 & 0  & 2  & 0  & 18  & 5.7k  & 2009 &  1087K & $\checkmark$ & $\checkmark$             \\
MonetDB    &   36  & 36  & 0 & 0 & 22  & 14  & 148 & 0.3k  & 2004 & 411K & $\checkmark$  & $\checkmark$        \\
MySQL      &  2   &  0  & 2 & 0  & 2  & 0  & 2   & 10.2k  & 1995 & 5532K& $\checkmark$    &   $\checkmark$   \\
Oracle     &  1   & 0 &  1  & 0  & 1 & 0     & 1 & - & 1977   & -    & -            & $\checkmark$ \\
Percona MySQL &  2   &  0  & 2 & 0  & 2  & 0  & 121 & 1.1k  & 2008 & 4182K  & -       & $\checkmark$       \\
RisingWave &  4   &  4  & 0 & 0  & 3  & 1  & 245 & 6.2k  & 2022 & 624K  & -          & -       \\
SQLite     &  3   &  3 & 0 & 0  & 3  & 0  & 10  & 5.4k  & 2000 & 372K  & $\checkmark$   & $\checkmark$  \\
TiDB       &  3   &  1  & 2 & 0  & 2  & 1  & 72  & 36.1k & 2016 & 1398K & $\checkmark$      & -    \\
Umbra      &  47  &  46 & 0 & 1 & 31 & 16 & -   & -     & 2018 &  -  & -     & $\checkmark$           \\
Virtuoso   &  10  &  10  & 0 & 0  & 8  & 2  & 83  & 0.8k  & 1998 & 2659K & -     & $\checkmark$          \\
Vitess     &  2   &  2  & 0 & 0  & 2  & 0  & 200  & 18.5k  & 2011 & 1533K & -     & -          \\
\midrule
\textbf{Total}  & \OverallBugs{} &\FixedBugs{} & \ConfirmedBugs{} & \DuplicatedBugs{}  & \LogicBugs{} & \OtherBugs{} & &  &   & & &                                      \\
\bottomrule
\end{tabular}

\end{table*}

\paragraph{Bug statistics}
Table~\ref{table:DBMSs} provides detailed statistics on the bugs we reported.
In total, we created \OverallBugs{} bug reports, and \FixedBugs{} bugs have been fixed as a direct response to our bug reports, which demonstrates that the DBMS developers considered most of the bugs important. 
Note that \LogicBugs{} of the bugs are logic bugs, found by the TLP~\cite{rigger2020finding} and NoREC~\cite{rigger2020detecting} oracles, with \TLPBugs{} and \NoRECBugs{} found respectively.
We found more bugs using TLP, because we first implemented and used it during testing.
In comparison, the original TLP and NoREC papers reported 77 and 51 logic bugs.
Our general \ProjectName{} showed promising ability in finding bugs, even though the generator was not specifically developed for the DBMS under test.
Although we tested the latest development versions, many of the bugs we found also affected the stable release version at the time of testing, such as all SQLite bugs and eight out of ten DuckDB bugs.
We reported only \DuplicatedBugs{} duplicate bugs, which shows the effectiveness of our bug prioritization approach, that we were careful with avoiding duplicate issues, and closely worked with the DBMS developers.

\paragraph{Developer reception}
Developer feedback is an important indicator of whether the bugs reported matter to developers. Our efforts received encouraging feedback.
For example, developers from CrateDB were curious about our method and wished to integrate our tool into their development cycle: \emph{``Yes, those bug reports are very helpful. We are interested in an automatic fuzz testing tool and we would integrate it into our development cycle. Please keep us updated!''}\footnote{\url{https://github.com/crate/crate/issues/15083}}
Developers from Dolt repeatedly expressed their gratitude, responding with comments such as: \emph{``This is very interesting. Keep the bugs coming. They are awesome.''}\footnote{\url{https://github.com/dolthub/dolt/issues/7018}}
Many of our reports were confirmed using replies such as \emph{``Thank you for the report''}, from MonetDB\footnote{\url{https://github.com/MonetDB/MonetDB/issues/7429}} and MySQL.\footnote{\url{https://bugs.mysql.com/bug.php?id=113298}}
We also reported bugs to CedarDB and Vitess, and their developers were highly receptive, expressing interest in adopting our tool to detect bugs.
The developers' reception shows the practical impact of our tool.
To illustrate this, we discuss multiple selected bugs below.

\begin{lstlisting}[language=SQL,float=tb,  caption={A bug in the \texttt{REPLACE} function remained undetected in SQLite for 10 years.}, label={listing:sqlite-bug}, escapechar=@, showspaces=false,
numbers=left,
xleftmargin=1em,
xrightmargin=-0.5cm,
numberstyle=\scriptsize,
commentstyle=\color{gray},
keywordstyle=\color{black}\bfseries,
breaklines=true,
basicstyle=\ttfamily\footnotesize,
breakatwhitespace]
CREATE TABLE t0(c0 TEXT, PRIMARY KEY(c0));
INSERT INTO t0 (c0) VALUES (1);

SELECT * FROM t0 WHERE t0.c0=REPLACE(1, '', 0);
-- 1 @\faCheck@
SELECT * FROM t0 
    WHERE NOT t0.c0=REPLACE(1, '', 0);
-- 1 @\faBug@
\end{lstlisting}

\begin{lstlisting}[language=SQL,float=tb,  caption={A bug in SQLite when handling multiple subqueries.}, label={listing:subquery-bug}, escapechar=@,
showspaces=false,
numbers=left,
xleftmargin=1em,
xrightmargin=-0.5cm,
numberstyle=\scriptsize,
commentstyle=\color{gray},
keywordstyle=\color{black}\bfseries,
breaklines=true,
basicstyle=\ttfamily\footnotesize,
breakatwhitespace]
CREATE TABLE t0(c0 INT);
CREATE TABLE t1(c0 INT);
INSERT INTO t0 (c0) VALUES (1);
CREATE VIEW v0(c0) AS
  SELECT 0 FROM t1 RIGHT JOIN t0 ON 1;

SELECT t0.c0
  FROM v0 LEFT JOIN (
    SELECT 'a' AS col0 FROM v0 WHERE false
  ) AS sub0 ON v0.c0,
    t0 RIGHT JOIN (
      SELECT NULL AS col0 FROM v0
  ) AS sub1 ON t0.c0; -- 1 @\faCheck@

SELECT t0.c0
  FROM v0 LEFT JOIN (
    SELECT 'a' AS col0 FROM v0 WHERE false
  ) AS sub0 ON v0.c0,
    t0 RIGHT JOIN (
      SELECT NULL AS col0 FROM v0
  ) AS sub1 ON t0.c0 WHERE t0.c0; -- {} @\faBug@
\end{lstlisting}

\paragraph{SQLite bug in \texttt{REPLACE} function}
Listing~\ref{listing:sqlite-bug} shows a bug in the implementation of the \texttt{REPLACE} function, which returned the first argument not a string but an intermediate object due to faulty implementation in SQLite, causing subsequent wrong comparison with the text. 
We found this bug using the TLP~\cite{rigger2020finding} oracle. The original query without a \texttt{WHERE} filter returns exactly one row in the table; however, the derived partitioned query, whose results should be composed to form the same result, returns two rows. The reason is that the query with the predicate and its negation both returns one row.
The fix was twofold: a direct fix to ensure the \texttt{REPLACE} function returns a \texttt{TEXT} value, and a deeper fix to make the comparison operator work normally even when one operand has both text and numeric types, which is a problem that could have been bisected to a commit in 2014.
This bug has been hidden for about ten years. Surprisingly, \ProjectName{} found it despite the efforts of \SQLancer{} and other logic bug detection tools.

\paragraph{SQLite bug with subquery}
Listing~\ref{listing:subquery-bug} shows another bug we found in SQLite, also detected by the TLP oracle. 
The second query should return one row since the predicate \texttt{t0.c0} should be evaluated to \texttt{1}.
The root cause of the bug is that the query flattener incorrectly changed an \texttt{ON} clause term to a \texttt{WHERE} clause term.
\SQLancer{} failed to detect this bug since it did not support subqueries, which is a feature newly integrated into \ProjectName{}.
Surprisingly, according to the developers' commit log,\footnote{\url{https://www.sqlite.org/src/info/bdd408a2557ff05c}} we found this is a follow-up bug of one bug that EET~\cite{jiang2024detecting} oracle detected a year ago.

\begin{figure}
    \centering
    \includegraphics[width=\linewidth]{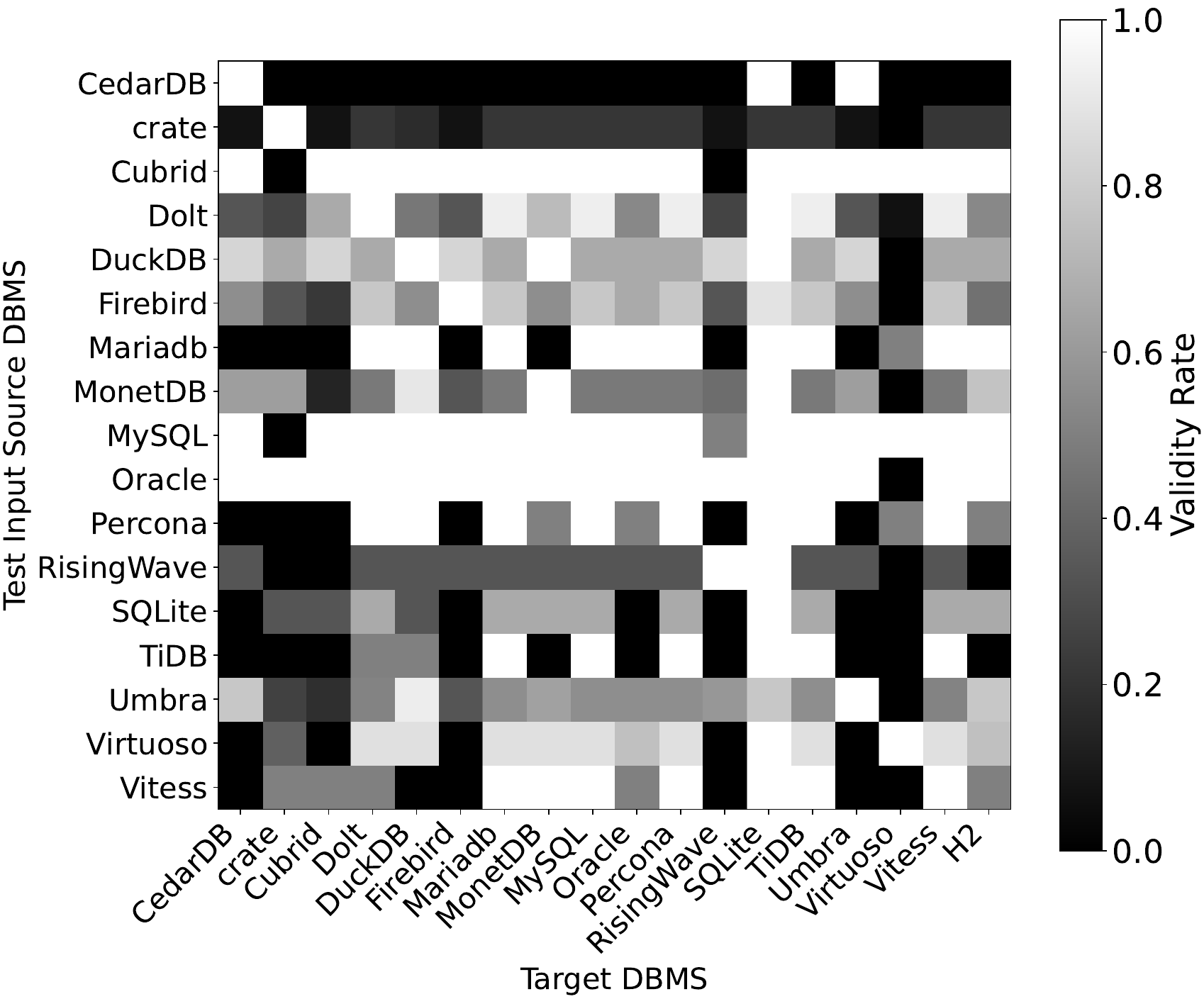}
    \caption{The validity rate of executing bug-inducing test cases across different DBMSs. At each intersection of \emph{bug source} and \emph{target DBMS}, the color represents the average success rate for executing bugs found in \emph{bug source} on the \emph{target DBMS}.}
    \label{fig:featureheatmap}
\end{figure}
\subsection{SQL Feature Study}\label{sec:featurestudy}
We conducted a case study on the \LogicBugs{} logic bugs that we found across different DBMSs to test our hypothesis that even for test cases with mostly common SQL features, still only a fraction of these features are supported by most DBMSs.
Intuitively, if we generated only the most common features, we would expect most DBMSs to support them all; however, our experiment results show that most of the bug-inducing features are unsupported on more than half of the DBMSs.
We considered only logic bugs since they returned incorrect results while executing without errors on the source DBMS. 
The H2 bug is excluded, since the developers fixed it by throwing an exception.
We took bug-inducing test cases found on a specific DBMS and executed them on all target DBMSs. 
Bug-inducing statements that ran without error were marked as successful; otherwise, they were marked as unsuccessful, indicating a syntax or semantic error due to the feature not being supported.
\paragraph{Result}
Figure~\ref{fig:featureheatmap} demonstrates the average validity rates when executing bug-inducing statements from each \emph{bug source} DBMS on \emph{target} DBMSs.
We found that \emph{none} of the bug-inducing tests can be executed successfully on \emph{all} 18 DBMSs. 
Only \SuccessfulBugsForMostDBMS{} of the bug-inducing test cases can be executed successfully on more than 90\% (17 of 18) DBMSs.
The overall valid rate of execution of bug-inducing test cases in different DBMSs is \OverallBugValidRate{}, indicating that features across DBMSs are mostly distinct and that generating these features is important for finding bugs.
These results demonstrate that most of the bugs cannot be executed successfully on other DBMSs, indicating that even though we generated mostly common features, they cannot be simply applied across all systems due to dialect deviations.
Only \ColumnsContainInvalidCases{} of \DBMSUnderTestNum{} DBMSs can execute test cases successfully from more than half of the source DBMSs, showing that it is difficult to directly use existing testing generators and apply them to other DBMSs.
The only one is SQLite which has a flexible type system, enabling most of the tests to be executed successfully.

\subsection{Bug Finding Efficiency}\label{sec: bug-efficiency}
We compared the efficiency in terms of bugs, unique query plan coverage and code coverage using multiple configurations of \ProjectName{} and \SQLancer{}.
First, we used the TLP oracle and evaluated \ProjectName{} with and without the validity feedback mechanism
on CrateDB for 1 hour across five runs.
We inspected the bugs found in one hour after prioritization.
Second, we evaluated the line and branch coverage of \ProjectName{} under different mechanisms and \SQLancer{} on three C/C++-based DBMSs, SQLite, PostgreSQL, and DuckDB. Code coverage is a common metric for fuzzing that assesses how much of a system might be tested and gives some indication of the features we covered. 
We observed that fuzzers for DBMSs can achieve higher code coverage than SQLancer and SQLancer++ even when executing only their seed corpus.

\paragraph{Bugs detection}
Table~\ref{tab:deduplication} shows the total number of bugs found within one hour.
This result shows that \ProjectName{} could detect different unique logic bugs under different random seeds efficiently in a limited time.
We observed that enabling the feedback mechanism resulted in more bugs being detected over the same period, indicating that the feedback mechanism can enhance bug-finding efficiency by increasing the validity rate of the generated queries.

\paragraph{Code coverage}
Table~\ref{tab:coverage} shows the average percentage of code coverage across 10 runs in 24 hours.
\SQLancer{} achieves higher code coverage in all three DBMSs. 
This is because \SQLancer{} generators are manually written for the specific DBMS and thus incorporate both features and heuristics specific to that DBMS, while \ProjectName{} supports only basic SQL statements and features; however, \ProjectName{} found 10 and 3 new bugs that were missed by \emph{SQLancer} in DuckDB and SQLite respectively (see Table~\ref{table:DBMSs}), demonstrating that the code coverage does not strongly relate to finding logic bugs. %
Typically, mutation-based coverage-guided fuzzers achieve higher coverage; for example, \emph{SQLRight} outperformed \emph{SQLancer} by 2\% and 39\% in line coverage for SQLite and PostgreSQL.
We also found that \SQLancer{} outperformed \ProjectName{} by 59\% and 21\% in line coverage for SQLite and PostgreSQL, likely due to developers putting more effort into making the generator more comprehensive; however, in DuckDB---where the generator might not be as mature---the gap was only 6\%. 
This demonstrates that our tool is especially beneficial for smaller developer teams, which, even if they adopt SQLancer, may be unable to incorporate many DBMS-specific features. For example, although MonetDB adopted SQLancer,\footnote{\url{https://github.com/MonetDB/sqlancer}} we still successfully uncovered 36 unique unknown bugs.

\begin{table}
    \centering
\footnotesize
\setlength{\tabcolsep}{4pt}
\caption{Coverage of \ProjectName{} and \SQLancer{} on SQLite, PostgreSQL and DuckDB in 24 hours}
\label{tab:coverage}
\begin{tabular}{lrrrrrr}
\toprule
\multirow{2}{*}{Approach}       & \multicolumn{2}{c}{SQLite}&  \multicolumn{2}{c}{PostgreSQL}&  \multicolumn{2}{c}{DuckDB}  \\
                                & Line  & Branch & Line  & Branch   & Line  & Branch    \\
\midrule
\emph{SQLancer++}         &  30.5\%   &  23.2\% &  26.3\% &  18.5\%  &  31.6\% & 18.8\% \\
\emph{SQLancer++\textsubscript{Rand}}        &  30.0\%   &   22.8\% &  26.1\% &  18.5\%  &  31.4\% &  18.5\% \\
\textit{SQLancer}    & 46.6\%  & 36.4\% & 32.3\% &  23.3\%  &  33.4\% & 20.9\%\\
\textit{SQLRight}          &  47.5\%   &  37.7\% & 44.9\% &  34.2\%  &   - & - \\

\bottomrule
\end{tabular}

\end{table}

\subsection{Feedback Mechanism}\label{sec: ablation-study}
We further evaluated the impact of validity feedback by measuring the query validity rates.
\paragraph{Methodology}
We ran \ProjectName{} on PostgreSQL and SQLite and measured the validity rate of generated SQL statements using different feedback mechanisms for 24 hours across 10 runs.
We calculated the validity rate by dividing the number of successfully executed test cases, each of which has two equivalent queries when using TLP oracle, by the total number of test cases.
The validity rate converged in less than one minute due to the high throughput of these two DBMSs; therefore, we did not measure the efficiency of the feedback mechanism.

\begin{table}[tb]
    \centering
\small
\caption{The validity rate of \ProjectName{} and \SQLancer{} for all queries on SQLite, PostgreSQL, and DuckDB over 24 hours.}
\label{tab:validity}
\begin{tabular}{lrrr}
\toprule
Approach    & SQLite & PostgreSQL & DuckDB \\
\midrule
\emph{SQLancer++}           & \SQLiteValidRateWithFB{} & \PostgreSQLValidRateWithFB{} & 64.2\%\\
\emph{SQLancer++\textsubscript{Rand}}         & \SQLiteValidRateWOFB{} & \PostgreSQLValidRateWOFB{}  & 24.6\%\\
\SQLancer{}                         & \SQLiteValidRateSQLancer{} &  \PostgreSQLValidRateSQLancer{} & 35.5\% \\
\bottomrule
\end{tabular}

\end{table}
\paragraph{Result}
Table~\ref{tab:validity} shows the average validity rate across 10 runs after 24 hours when executing \ProjectName{} and \SQLancer{} on SQLite,  PostgreSQL, and DuckDB.
In general, we observed a higher validity rate for SQLite due to its dynamic type system; conversely, PostgreSQL frequently raised errors for type mismatches and other errors and we thus observed a lower success rate.
For SQLite, the validity rate was \SQLiteValidRateWithFB{} when enabling feedback, increasing \SQLiteValidRateIncrease{} compared with not enabling it. For PostgreSQL, the validity rate was \PostgreSQLValidRateWithFB{}, increasing \PostgreSQLValideRateIncrease{}. 
The rather lower validity rate of the \SQLancer{} PostgreSQL generator might be due to its incorporation of additional DBMS-specific features, leading to a higher failure rate due to their complexity.

\subsection{Bug Prioritization Effectiveness}\label{sec: bug-deduplication}
To evaluate the effectiveness of our bug-prioritization approach, we determined how well it could identify bug-inducing test cases that triggered unique bugs. %

\paragraph{Methodology}
We conducted a case study on an open-source DBMS, CrateDB, from which we could obtain the commit logs. 
This enabled us to distinguish bugs by reproducing them on different versions of the system to see if they have been fixed by distinct commits.
We used the TLP oracle, and enabled the feedback mechanism. 
We evaluated \ProjectName{} on CrateDB for one hour of continuous running five times, recording any bugs that occurred. 
We ran the test for only one hour because CrateDB would run out of memory when we ran for multiple hours.
Within the one-hour period, \ProjectName{} could still trigger numerous bugs.
We analyzed the unique bugs identified by \ProjectName{} by bisecting them across different versions of the system. %

\begin{table}[tb]
    \centering
\small
\caption{The average number of all, prioritized and unique bugs found in CrateDB across 5 runs in 1 hour}
\label{tab:deduplication}
\begin{tabular}{lrrr}
\toprule
\multirow{2}{*}{Approach}    & Detected & Prioritized & Unique \\
                            & Bugs     & Bugs         & Bugs   \\
\midrule

\emph{SQLancer++}          & 67,878.2 & \AvgCrateDBBugswithFB{} &  \AvgCrateDBUniqueBugswithFB{}  \\
\emph{SQLancer++\textsubscript{Rand}}         & 55,412.2 &\AvgCrateDBBugswoFB{} & \AvgCrateDBUniqueBugswoFB{}  \\
\bottomrule

\end{tabular}

\end{table}

\paragraph{Result}
Table~\ref{tab:deduplication} shows the results of bugs found by \ProjectName{} in one hour across five test runs.
\ProjectName{} could detect more than 60K bug-inducing test cases. 
\ProjectName{} identified bugs that were potentially caused by the same reason and reported only \AvgCrateDBBugswithFB{} and \AvgCrateDBBugswoFB{} bugs, reducing more than 99\% of the duplicated bugs.
On average, among the bugs detected and prioritized with and without feedback mechanism, \AvgCrateDBUniqueBugswithFB{} and \AvgCrateDBUniqueBugswoFB{} bugs were unique respectively. 
More than half of the bugs were duplicated since bug-inducing test cases with different features may be due to the same reasons (\emph{e.g.}, unequal operator \texttt{``<>''} and \texttt{``!=''}).
Despite this, the results show the effectiveness of prioritization, which could help developers prioritize the bugs.

\section{Discussion}
\paragraph{Found bugs}
As shown in Table~\ref{table:DBMSs}, \LogicBugs{} of \OverallBugs{} bugs were logic bugs, which shows the effectiveness of our approach. 
Among the \OtherBugs{} other bugs, \PerformanceBugs{} of them were performance issues, \ErrorBugs{} of them were due to unexpected errors (\emph{e.g.}, connection failures), and \CrashBugs{} of them were crash bugs. 
We believe logic bugs are more severe and harder to detect than other kinds of bugs; crashes, and internal errors are immediately visible to users and developers, while logic bugs can go unnoticed.
Furthermore, our current implementation includes mostly commonly used features, meaning that bugs could impact users relying on these features.
These reasons could explain the high fix rate for our reported bugs.

\paragraph{Comparison with \SQLancer{}}
In addition to the bugs \ProjectName{} found in the previously untested DBMSs, it also found previously unknown bugs in the well-tested DBMSs, for example, MySQL and SQLite.
Of the \OverallBugs{} bugs, 138 occurred in DBMSs not supported by \SQLancer{}, while the remaining 58 occurred in DBMSs that \SQLancer{} supported, but relied on features unsupported by its current generators.
For example, \SQLancer{} lacks the scalar function \texttt{REPLACE}, for which we found a related bug in SQLite (see Listing~\ref{listing:sqlite-bug}).
Besides, \SQLancer{} supports all bit operators except for the bitwise inversion (\texttt{\textasciitilde}) in TiDB, in which we discovered a bug.\footnote{\url{https://github.com/pingcap/tidb/issues/53506}}
In general, we expect \SQLancer{} to find more bugs than \ProjectName{}, as \SQLancer{} uses the same test oracles for finding logic bugs, but provides manually-written generators for each DBMS it supports.
\ProjectName{} and \SQLancer{} are thus complementary: \ProjectName{} scales to many DBMSs with minimal effort, whereas \SQLancer{} can provide deeper testing for individual systems with manually written generators.

\paragraph{Manual effort}
Limited manual effort is still required to apply \ProjectName{} to new DBMSs.
First, it is necessary to manually include the database driver and specify the connection string, indicating the port, user, and password. 
Second, for some DBMSs, after schema generation, \ProjectName{} must ensure that the data inserted can be read. 
SQL statements like \texttt{COMMIT} (implicitly specified by JDBC) or customized statements for distributed systems might need to be used, for example, the \texttt{REFRESH} statement in CrateDB. 
These are explicitly issued for DBMSs that require them.
These adaptation efforts are minor, as we implemented support for \SupportedDBMSs{} DBMSs with each \AvgLOCDBMS{} LOC per DBMS on average. 
For most DBMSs, only 4 LOC are required to specify the connection string.
Finally, we also created Docker containers for each system under test, requiring additional effort.
The automatic generation of fuzz drivers is an active research area~\cite{Ispoglou2020FuzzGen, babic2019fudge, zhang2023understanding}, and we believe that some of their insights could also help the automation of this functionality in DBMS testing.

\section{Related Work}

\paragraph{Testing platforms}
Two other large-scale deployments of bug-finding platforms have inspired our work on \ProjectName{}, namely OSS-Fuzz~\cite{kostya2017oss} and syzkaller~\cite{syzkaller}. %
OSS-Fuzz supports common fuzzers such as libfuzz~\cite{libfuzzer}, AFL++~\cite{fioraldi2020afl++}, Honggfuzz~\cite{honggfuzz}, and ClusterFuzz~\cite{clusterfuzz}.
These fuzzers are grey-box fuzzers, which use code coverage feedback from the tested program to prioritize which inputs should be further mutated.
Unlike \ProjectName{}, OSS-Fuzz can apply these fuzzers to any program for which a fuzz harness is written, but typically cannot find logic bugs, as the fuzzers rely on crashes or sanitizers~\cite{konstantin2012addresssanitizer, stepanov2015memorysanitizer, serebryany2009threadsanitizer}, the latter which detect undefined behavior in C/C++, as test oracles.
Syzkaller is a coverage-guided kernel fuzzer, which finds bugs in operating systems by mutating sequences of system calls.
As the above fuzzers, syzkaller aims to find crash bugs.
In contrast to OSS-Fuzz and syzkaller, the key challenge in our context is that hundreds of DBMSs with different SQL dialects exist, which we aim to support.
A previous study~\cite{liang2024wingfuzz} investigated the challenges of applying fuzzing tools for DBMSs in a continuous integration process, while it focused on crash bugs, it highlighted related challenges such as those we tackled.
We aim to find deeper kinds of bugs, such as logic bugs.

\paragraph{DBMS test oracles}
\ProjectName{} can use any test oracle that is not based on DBMS-specific features.
In this work, we have used TLP and NoREC.
TLP~\cite{rigger2020finding} finds logic bugs by comparing an original query with an equivalent query derived from it by partitioning the query's results into three parts. NoREC~\cite{rigger2020detecting} detects logic bugs by comparing the results of a query that is receptive to optimizations with an equivalent one that the DBMS is likely to fail to optimize.
Various other test oracles could be adopted.
Differential Query Execution (DQE)~\cite{song2023testing} uses the same predicate in different SQL statements, assuming that the predicate consistently evaluates to the same value for a given statement.
Pinolo~\cite{hao2023pinolo} generates pairs of queries that are in superset or subset relations and checks whether the expected relation is met.
EET~\cite{jiang2024detecting} transforms given queries and checks whether the transformed versions still produce the same results as the original query.
However, several oracles might be difficult to support in \ProjectName{} due to dependencies on DBMS-specific features. For example, CERT~\cite{ba2023finding} requires parsing query plans, which are typically DBMS-specific, to find unexpected discrepancies in the DBMSs' cardinality estimator.
DQP~\cite{ba2024keep} detects logic bugs by using DBMS-specific query hints, and Mozi~\cite{mozi} through DBMS-specific configurations, neither of which could be easily supported.

\paragraph{Test-case generation for DBMSs}
Various approaches have been proposed to automatically generate test inputs for DBMSs.
The most closely related work is Griffin~\cite{fu2023griffin}, which was proposed as an alternative to grammar-based fuzzers by maintaining lightweight metadata related to the DBMS state to improve mutation correctness in fuzzing.
While this enables it to fuzz a range of DBMSs, as a mutation-based fuzzer based on AFL++, it requires a diverse seed input corpus and aims to find only crash bugs, which limits the scope and scalability.
BuzzBee~\cite{yang2024towards} aims to fuzz various kinds of DBMSs, which include NoSQL systems. However, it shares a similar limitation with Griffin that it cannot be applied to find logic bugs.
DynSQL~\cite{jiang2023dynsql} incorporates error feedback by the DBMSs to incrementally expand a given SQL query while maintaining its validity to find crash bugs in DBMSs.
In contrast to DynSQL, our feedback applies to potentially unknown DBMSs and to entire SQL statements aiming to find logic bugs.
SQLRight~\cite{liang2022detecting}, inspired by grey-box fuzzers, uses code-coverage feedback in combination with test oracles such as NoREC and TLP to find bugs.
Query Plan Guidance (QPG)~\cite{ba2023testing} uses query plans as a feedback mechanism to determine whether a given database state has saturated for finding potential bugs.
Squirrel~\cite{zhong2020squirrel} is a mutation-based method to generate new queries for finding memory errors.
None of the above approaches explored generators that could apply to the multitude of existing DBMSs aiming to find logic bugs.

\section{Conclusion}
In this paper, we have presented a simple and effective adaptive query generator, which is at the core of a new large-scale automated testing platform for DBMSs, called \ProjectName{}, which also includes new techniques to prioritize bugs as well as model the schema of the DBMS under test.
As part of our initial efforts, \ProjectName{} has enabled developers of \DBMSUnderTestNum{} DBMSs to fix \FixedBugs{} previously unknown bugs in their systems.
The adaptive query generator is only a first step towards our vision of fully automated DBMS testing. Various challenges remain to be tackled, such as covering uncommon features in the generator, designing more sophisticated policies to utilize DBMS feedback, or prioritizing bug reports.
We hope that \ProjectName{} will eventually become a standard tool in DBMS developers' toolboxes.

\begin{acks}
We would like to sincerely thank all the anonymous reviewers and our shepherd, Pedro Fonseca, for their valuable feedback and insights that helped us improve the quality of this paper. 
We want to thank all the DBMS developers for responding to our bug reports and for analyzing and fixing the issues we identified.
This research is supported by an Amazon Research Award Fall 2023 as well as the National Research Foundation, Singapore, and Cyber Security Agency of Singapore under its National Cybersecurity R\&D Programme (Fuzz Testing). Any opinions, findings and conclusions, or recommendations expressed in this material are those of the author(s) and do not reflect the views of Amazon, National Research Foundation, Singapore, and Cyber Security Agency of Singapore.
\end{acks}

\bibliographystyle{ACM-Reference-Format}
\balance
\bibliography{sample-base}


\begin{thebibliography}{54}


\ifx \showCODEN    \undefined \def \showCODEN     #1{\unskip}     \fi
\ifx \showISBNx    \undefined \def \showISBNx     #1{\unskip}     \fi
\ifx \showISBNxiii \undefined \def \showISBNxiii  #1{\unskip}     \fi
\ifx \showISSN     \undefined \def \showISSN      #1{\unskip}     \fi
\ifx \showLCCN     \undefined \def \showLCCN      #1{\unskip}     \fi
\ifx \shownote     \undefined \def \shownote      #1{#1}          \fi
\ifx \showarticletitle \undefined \def \showarticletitle #1{#1}   \fi
\ifx \showURL      \undefined \def \showURL       {\relax}        \fi
\providecommand\bibfield[2]{#2}
\providecommand\bibinfo[2]{#2}
\providecommand\natexlab[1]{#1}
\providecommand\showeprint[2][]{arXiv:#2}

\bibitem[jaz(2021)]%
        {jazzer}
 \bibinfo{year}{2021}\natexlab{}.
\newblock \bibinfo{title}{Jazzer: Coverage-guided, in-process fuzzing for the JVM}.
\newblock \bibinfo{howpublished}{https://github.com/CodeIntelligenceTesting/jazzer}.
\newblock


\bibitem[Ba and Rigger(2023)]%
        {ba2023testing}
\bibfield{author}{\bibinfo{person}{Jinsheng Ba} {and} \bibinfo{person}{Manuel Rigger}.} \bibinfo{year}{2023}\natexlab{}.
\newblock \showarticletitle{Testing Database Engines via Query Plan Guidance}. In \bibinfo{booktitle}{\emph{Proceedings of the 45th International Conference on Software Engineering}} (Melbourne, Victoria, Australia) \emph{(\bibinfo{series}{ICSE '23})}. \bibinfo{publisher}{IEEE Press}, \bibinfo{pages}{2060–2071}.
\newblock
\showISBNx{9781665457019}
\href{https://doi.org/10.1109/ICSE48619.2023.00174}{doi:\nolinkurl{10.1109/ICSE48619.2023.00174}}


\bibitem[Ba and Rigger(2024a)]%
        {ba2023finding}
\bibfield{author}{\bibinfo{person}{Jinsheng Ba} {and} \bibinfo{person}{Manuel Rigger}.} \bibinfo{year}{2024}\natexlab{a}.
\newblock \showarticletitle{CERT: Finding Performance Issues in Database Systems Through the Lens of Cardinality Estimation}. In \bibinfo{booktitle}{\emph{Proceedings of the IEEE/ACM 46th International Conference on Software Engineering}} (, Lisbon, Portugal,) \emph{(\bibinfo{series}{ICSE '24})}. \bibinfo{publisher}{Association for Computing Machinery}, \bibinfo{address}{New York, NY, USA}, Article \bibinfo{articleno}{133}, \bibinfo{numpages}{13}~pages.
\newblock
\showISBNx{9798400702174}
\href{https://doi.org/10.1145/3597503.3639076}{doi:\nolinkurl{10.1145/3597503.3639076}}


\bibitem[Ba and Rigger(2024b)]%
        {ba2024keep}
\bibfield{author}{\bibinfo{person}{Jinsheng Ba} {and} \bibinfo{person}{Manuel Rigger}.} \bibinfo{year}{2024}\natexlab{b}.
\newblock \showarticletitle{Keep It Simple: Testing Databases via Differential Query Plans}.
\newblock \bibinfo{journal}{\emph{Proc. ACM Manag. Data}} \bibinfo{volume}{2}, \bibinfo{number}{3}, Article \bibinfo{articleno}{188} (\bibinfo{date}{May} \bibinfo{year}{2024}), \bibinfo{numpages}{26}~pages.
\newblock
\href{https://doi.org/10.1145/3654991}{doi:\nolinkurl{10.1145/3654991}}


\bibitem[Babi\'{c} et~al\mbox{.}(2019)]%
        {babic2019fudge}
\bibfield{author}{\bibinfo{person}{Domagoj Babi\'{c}}, \bibinfo{person}{Stefan Bucur}, \bibinfo{person}{Yaohui Chen}, \bibinfo{person}{Franjo Ivan\v{c}i\'{c}}, \bibinfo{person}{Tim King}, \bibinfo{person}{Markus Kusano}, \bibinfo{person}{Caroline Lemieux}, \bibinfo{person}{L\'{a}szl\'{o} Szekeres}, {and} \bibinfo{person}{Wei Wang}.} \bibinfo{year}{2019}\natexlab{}.
\newblock \showarticletitle{FUDGE: fuzz driver generation at scale}. In \bibinfo{booktitle}{\emph{Proceedings of the 2019 27th ACM Joint Meeting on European Software Engineering Conference and Symposium on the Foundations of Software Engineering}} (Tallinn, Estonia) \emph{(\bibinfo{series}{ESEC/FSE 2019})}. \bibinfo{publisher}{Association for Computing Machinery}, \bibinfo{address}{New York, NY, USA}, \bibinfo{pages}{975–985}.
\newblock
\showISBNx{9781450355728}
\href{https://doi.org/10.1145/3338906.3340456}{doi:\nolinkurl{10.1145/3338906.3340456}}


\bibitem[Chen et~al\mbox{.}(2013)]%
        {chen2013taming}
\bibfield{author}{\bibinfo{person}{Yang Chen}, \bibinfo{person}{Alex Groce}, \bibinfo{person}{Chaoqiang Zhang}, \bibinfo{person}{Weng-Keen Wong}, \bibinfo{person}{Xiaoli Fern}, \bibinfo{person}{Eric Eide}, {and} \bibinfo{person}{John Regehr}.} \bibinfo{year}{2013}\natexlab{}.
\newblock \showarticletitle{Taming compiler fuzzers}. In \bibinfo{booktitle}{\emph{Proceedings of the 34th ACM SIGPLAN Conference on Programming Language Design and Implementation}} (Seattle, Washington, USA) \emph{(\bibinfo{series}{PLDI '13})}. \bibinfo{publisher}{Association for Computing Machinery}, \bibinfo{address}{New York, NY, USA}, \bibinfo{pages}{197–208}.
\newblock
\showISBNx{9781450320146}
\href{https://doi.org/10.1145/2491956.2462173}{doi:\nolinkurl{10.1145/2491956.2462173}}


\bibitem[Company(2024)]%
        {DatabaseSoftwareMarket2024}
\bibfield{author}{\bibinfo{person}{The Business~Research Company}.} \bibinfo{year}{2024}\natexlab{}.
\newblock \bibinfo{booktitle}{\emph{Database Software Global Market Report 2024}}.
\newblock \bibinfo{publisher}{{Research and Markets}}.
\newblock
\urldef\tempurl%
\url{https://www.researchandmarkets.com/report/database-software}
\showURL{%
\tempurl}


\bibitem[Fioraldi et~al\mbox{.}(2020)]%
        {fioraldi2020afl++}
\bibfield{author}{\bibinfo{person}{Andrea Fioraldi}, \bibinfo{person}{Dominik Maier}, \bibinfo{person}{Heiko Ei{\ss}feldt}, {and} \bibinfo{person}{Marc Heuse}.} \bibinfo{year}{2020}\natexlab{}.
\newblock \showarticletitle{$\{$AFL++$\}$: Combining incremental steps of fuzzing research}. In \bibinfo{booktitle}{\emph{14th USENIX Workshop on Offensive Technologies (WOOT 20)}}.
\newblock


\bibitem[Fu et~al\mbox{.}(2023)]%
        {fu2023griffin}
\bibfield{author}{\bibinfo{person}{Jingzhou Fu}, \bibinfo{person}{Jie Liang}, \bibinfo{person}{Zhiyong Wu}, \bibinfo{person}{Mingzhe Wang}, {and} \bibinfo{person}{Yu Jiang}.} \bibinfo{year}{2023}\natexlab{}.
\newblock \showarticletitle{Griffin: Grammar-Free DBMS Fuzzing}. In \bibinfo{booktitle}{\emph{Proceedings of the 37th IEEE/ACM International Conference on Automated Software Engineering}} (Rochester, MI, USA) \emph{(\bibinfo{series}{ASE '22})}. \bibinfo{publisher}{Association for Computing Machinery}, \bibinfo{address}{New York, NY, USA}, Article \bibinfo{articleno}{49}, \bibinfo{numpages}{12}~pages.
\newblock
\showISBNx{9781450394758}
\href{https://doi.org/10.1145/3551349.3560431}{doi:\nolinkurl{10.1145/3551349.3560431}}


\bibitem[Gao et~al\mbox{.}(2023)]%
        {gao2023comprehensive}
\bibfield{author}{\bibinfo{person}{Xiyue Gao}, \bibinfo{person}{Zhuang Liu}, \bibinfo{person}{Jiangtao Cui}, \bibinfo{person}{Hui Li}, \bibinfo{person}{Hui Zhang}, \bibinfo{person}{Kewei Wei}, {and} \bibinfo{person}{Kankan Zhao}.} \bibinfo{year}{2023}\natexlab{}.
\newblock \bibinfo{title}{A Comprehensive Survey on Database Management System Fuzzing: Techniques, Taxonomy and Experimental Comparison}.
\newblock
\showeprint[arxiv]{2311.06728}~[cs.DB]


\bibitem[Gelman et~al\mbox{.}(1995)]%
        {gelman1995bayesian}
\bibfield{author}{\bibinfo{person}{Andrew Gelman}, \bibinfo{person}{John~B Carlin}, \bibinfo{person}{Hal~S Stern}, {and} \bibinfo{person}{Donald~B Rubin}.} \bibinfo{year}{1995}\natexlab{}.
\newblock \bibinfo{booktitle}{\emph{Bayesian data analysis}}.
\newblock \bibinfo{publisher}{Chapman and Hall/CRC}.
\newblock


\bibitem[{Google}(2016)]%
        {honggfuzz}
\bibfield{author}{\bibinfo{person}{{Google}}.} \bibinfo{year}{2016}\natexlab{}.
\newblock \bibinfo{title}{Honggfuzz}.
\newblock \bibinfo{howpublished}{\url{https://honggfuzz.dev/}}.
\newblock
\newblock
\shownote{Accessed: 2024-03-30}.


\bibitem[{Google}(2018)]%
        {clusterfuzz}
\bibfield{author}{\bibinfo{person}{{Google}}.} \bibinfo{year}{2018}\natexlab{}.
\newblock \bibinfo{title}{ClusterFuzz - Scalable Fuzzing Infrastructure}.
\newblock \bibinfo{howpublished}{\url{https://google.github.io/clusterfuzz/}}.
\newblock
\newblock
\shownote{Accessed: 2024-03-30}.


\bibitem[{Google}(2023)]%
        {syzkaller}
\bibfield{author}{\bibinfo{person}{{Google}}.} \bibinfo{year}{2023}\natexlab{}.
\newblock \bibinfo{title}{{syzkaller: unsupervised, coverage-guided kernel fuzzer}}.
\newblock \bibinfo{howpublished}{\url{https://github.com/google/syzkaller}}.
\newblock
\newblock
\shownote{Accessed: 2024-02-20}.


\bibitem[Hao et~al\mbox{.}(2023)]%
        {hao2023pinolo}
\bibfield{author}{\bibinfo{person}{Zongyin Hao}, \bibinfo{person}{Quanfeng Huang}, \bibinfo{person}{Chengpeng Wang}, \bibinfo{person}{Jianfeng Wang}, \bibinfo{person}{Yushan Zhang}, \bibinfo{person}{Rongxin Wu}, {and} \bibinfo{person}{Charles Zhang}.} \bibinfo{year}{2023}\natexlab{}.
\newblock \showarticletitle{Pinolo: Detecting Logical Bugs in Database Management Systems with Approximate Query Synthesis}. In \bibinfo{booktitle}{\emph{2023 USENIX Annual Technical Conference (USENIX ATC 23)}}. \bibinfo{pages}{345--358}.
\newblock


\bibitem[Hilprecht et~al\mbox{.}(2020)]%
        {hilprecht2020learning}
\bibfield{author}{\bibinfo{person}{Benjamin Hilprecht}, \bibinfo{person}{Carsten Binnig}, {and} \bibinfo{person}{Uwe R\"{o}hm}.} \bibinfo{year}{2020}\natexlab{}.
\newblock \showarticletitle{Learning a Partitioning Advisor for Cloud Databases}. In \bibinfo{booktitle}{\emph{Proceedings of the 2020 ACM SIGMOD International Conference on Management of Data}} (Portland, OR, USA) \emph{(\bibinfo{series}{SIGMOD '20})}. \bibinfo{publisher}{Association for Computing Machinery}, \bibinfo{address}{New York, NY, USA}, \bibinfo{pages}{143–157}.
\newblock
\showISBNx{9781450367356}
\href{https://doi.org/10.1145/3318464.3389704}{doi:\nolinkurl{10.1145/3318464.3389704}}


\bibitem[Huang et~al\mbox{.}(2019)]%
        {huang2019xengine}
\bibfield{author}{\bibinfo{person}{Gui Huang}, \bibinfo{person}{Xuntao Cheng}, \bibinfo{person}{Jianying Wang}, \bibinfo{person}{Yujie Wang}, \bibinfo{person}{Dengcheng He}, \bibinfo{person}{Tieying Zhang}, \bibinfo{person}{Feifei Li}, \bibinfo{person}{Sheng Wang}, \bibinfo{person}{Wei Cao}, {and} \bibinfo{person}{Qiang Li}.} \bibinfo{year}{2019}\natexlab{}.
\newblock \showarticletitle{X-Engine: An Optimized Storage Engine for Large-scale E-commerce Transaction Processing}. In \bibinfo{booktitle}{\emph{Proceedings of the 2019 International Conference on Management of Data}} (Amsterdam, Netherlands) \emph{(\bibinfo{series}{SIGMOD '19})}. \bibinfo{publisher}{Association for Computing Machinery}, \bibinfo{address}{New York, NY, USA}, \bibinfo{pages}{651–665}.
\newblock
\showISBNx{9781450356435}
\href{https://doi.org/10.1145/3299869.3314041}{doi:\nolinkurl{10.1145/3299869.3314041}}


\bibitem[Ispoglou et~al\mbox{.}(2020)]%
        {Ispoglou2020FuzzGen}
\bibfield{author}{\bibinfo{person}{Kyriakos~K. Ispoglou}, \bibinfo{person}{Daniel Austin}, \bibinfo{person}{Vishwath Mohan}, {and} \bibinfo{person}{Mathias Payer}.} \bibinfo{year}{2020}\natexlab{}.
\newblock \showarticletitle{FuzzGen: automatic fuzzer generation}. In \bibinfo{booktitle}{\emph{Proceedings of the 29th USENIX Conference on Security Symposium}} \emph{(\bibinfo{series}{SEC'20})}. \bibinfo{publisher}{USENIX Association}, \bibinfo{address}{USA}, Article \bibinfo{articleno}{128}, \bibinfo{numpages}{17}~pages.
\newblock
\showISBNx{978-1-939133-17-5}


\bibitem[Jiang et~al\mbox{.}(2021)]%
        {jiang2021igor}
\bibfield{author}{\bibinfo{person}{Zhiyuan Jiang}, \bibinfo{person}{Xiyue Jiang}, \bibinfo{person}{Ahmad Hazimeh}, \bibinfo{person}{Chaojing Tang}, \bibinfo{person}{Chao Zhang}, {and} \bibinfo{person}{Mathias Payer}.} \bibinfo{year}{2021}\natexlab{}.
\newblock \showarticletitle{Igor: Crash Deduplication Through Root-Cause Clustering}. In \bibinfo{booktitle}{\emph{Proceedings of the 2021 ACM SIGSAC Conference on Computer and Communications Security}} (Virtual Event, Republic of Korea) \emph{(\bibinfo{series}{CCS '21})}. \bibinfo{publisher}{Association for Computing Machinery}, \bibinfo{address}{New York, NY, USA}, \bibinfo{pages}{3318–3336}.
\newblock
\showISBNx{9781450384544}
\href{https://doi.org/10.1145/3460120.3485364}{doi:\nolinkurl{10.1145/3460120.3485364}}


\bibitem[Jiang et~al\mbox{.}(2023a)]%
        {jiang2023dynsql}
\bibfield{author}{\bibinfo{person}{Zu-Ming Jiang}, \bibinfo{person}{Jia-Ju Bai}, {and} \bibinfo{person}{Zhendong Su}.} \bibinfo{year}{2023}\natexlab{a}.
\newblock \showarticletitle{{DynSQL}: Stateful Fuzzing for Database Management Systems with Complex and Valid {SQL} Query Generation}. In \bibinfo{booktitle}{\emph{32nd USENIX Security Symposium (USENIX Security 23)}}. \bibinfo{publisher}{USENIX Association}, \bibinfo{address}{Anaheim, CA}, \bibinfo{pages}{4949--4965}.
\newblock
\showISBNx{978-1-939133-37-3}
\urldef\tempurl%
\url{https://www.usenix.org/conference/usenixsecurity23/presentation/jiang-zu-ming}
\showURL{%
\tempurl}


\bibitem[Jiang et~al\mbox{.}(2023b)]%
        {jiang2023detecting}
\bibfield{author}{\bibinfo{person}{Zu-Ming Jiang}, \bibinfo{person}{Si Liu}, \bibinfo{person}{Manuel Rigger}, {and} \bibinfo{person}{Zhendong Su}.} \bibinfo{year}{2023}\natexlab{b}.
\newblock \showarticletitle{Detecting Transactional Bugs in Database Engines via {Graph-Based} Oracle Construction}. In \bibinfo{booktitle}{\emph{17th USENIX Symposium on Operating Systems Design and Implementation (OSDI 23)}}. \bibinfo{publisher}{USENIX Association}, \bibinfo{address}{Boston, MA}, \bibinfo{pages}{397--417}.
\newblock
\showISBNx{978-1-939133-34-2}
\urldef\tempurl%
\url{https://www.usenix.org/conference/osdi23/presentation/jiang}
\showURL{%
\tempurl}


\bibitem[Jiang and Su(2024)]%
        {jiang2024detecting}
\bibfield{author}{\bibinfo{person}{Zu-Ming Jiang} {and} \bibinfo{person}{Zhendong Su}.} \bibinfo{year}{2024}\natexlab{}.
\newblock \showarticletitle{Detecting Logic Bugs in Database Engines via Equivalent Expression Transformation}. In \bibinfo{booktitle}{\emph{18th USENIX Symposium on Operating Systems Design and Implementation (OSDI 24)}}. \bibinfo{publisher}{USENIX Association}, \bibinfo{address}{Santa Clara, CA}, \bibinfo{pages}{821--835}.
\newblock
\showISBNx{978-1-939133-40-3}
\urldef\tempurl%
\url{https://www.usenix.org/conference/osdi24/presentation/jiang}
\showURL{%
\tempurl}


\bibitem[Korolija et~al\mbox{.}(2022)]%
        {korolija2022farview}
\bibfield{author}{\bibinfo{person}{Dario Korolija}, \bibinfo{person}{Dimitrios Koutsoukos}, \bibinfo{person}{Kimberly Keeton}, \bibinfo{person}{Konstantin Taranov}, \bibinfo{person}{Dejan~S. Milojicic}, {and} \bibinfo{person}{Gustavo Alonso}.} \bibinfo{year}{2022}\natexlab{}.
\newblock \showarticletitle{Farview: Disaggregated Memory with Operator Off-loading for Database Engines}. In \bibinfo{booktitle}{\emph{12th Conference on Innovative Data Systems Research, {CIDR} 2022, Chaminade, CA, USA, January 9-12, 2022}}. \bibinfo{publisher}{www.cidrdb.org}.
\newblock
\urldef\tempurl%
\url{https://www.cidrdb.org/cidr2022/papers/p11-korolija.pdf}
\showURL{%
\tempurl}


\bibitem[Kraska et~al\mbox{.}(2019)]%
        {kraska2019sagedb}
\bibfield{author}{\bibinfo{person}{Tim Kraska}, \bibinfo{person}{Mohammad Alizadeh}, \bibinfo{person}{Alex Beutel}, \bibinfo{person}{Ed~H. Chi}, \bibinfo{person}{Ani Kristo}, \bibinfo{person}{Guillaume Leclerc}, \bibinfo{person}{Samuel Madden}, \bibinfo{person}{Hongzi Mao}, {and} \bibinfo{person}{Vikram Nathan}.} \bibinfo{year}{2019}\natexlab{}.
\newblock \showarticletitle{SageDB: {A} Learned Database System}. In \bibinfo{booktitle}{\emph{9th Biennial Conference on Innovative Data Systems Research, {CIDR} 2019, Asilomar, CA, USA, January 13-16, 2019, Online Proceedings}}. \bibinfo{publisher}{www.cidrdb.org}.
\newblock
\urldef\tempurl%
\url{http://cidrdb.org/cidr2019/papers/p117-kraska-cidr19.pdf}
\showURL{%
\tempurl}


\bibitem[Le et~al\mbox{.}(2014)]%
        {le2014compiler}
\bibfield{author}{\bibinfo{person}{Vu Le}, \bibinfo{person}{Mehrdad Afshari}, {and} \bibinfo{person}{Zhendong Su}.} \bibinfo{year}{2014}\natexlab{}.
\newblock \showarticletitle{Compiler validation via equivalence modulo inputs}. In \bibinfo{booktitle}{\emph{Proceedings of the 35th ACM SIGPLAN Conference on Programming Language Design and Implementation}} (Edinburgh, United Kingdom) \emph{(\bibinfo{series}{PLDI '14})}. \bibinfo{publisher}{Association for Computing Machinery}, \bibinfo{address}{New York, NY, USA}, \bibinfo{pages}{216–226}.
\newblock
\showISBNx{9781450327848}
\href{https://doi.org/10.1145/2594291.2594334}{doi:\nolinkurl{10.1145/2594291.2594334}}


\bibitem[Lecoeur et~al\mbox{.}(2023)]%
        {lecoeur2023program}
\bibfield{author}{\bibinfo{person}{Bastien Lecoeur}, \bibinfo{person}{Hasan Mohsin}, {and} \bibinfo{person}{Alastair~F. Donaldson}.} \bibinfo{year}{2023}\natexlab{}.
\newblock \showarticletitle{Program Reconditioning: Avoiding Undefined Behaviour When Finding and Reducing Compiler Bugs}.
\newblock \bibinfo{journal}{\emph{Proc. ACM Program. Lang.}} \bibinfo{volume}{7}, \bibinfo{number}{PLDI}, Article \bibinfo{articleno}{180} (\bibinfo{date}{jun} \bibinfo{year}{2023}), \bibinfo{numpages}{25}~pages.
\newblock
\href{https://doi.org/10.1145/3591294}{doi:\nolinkurl{10.1145/3591294}}


\bibitem[Liang et~al\mbox{.}(2024a)]%
        {liang2024wingfuzz}
\bibfield{author}{\bibinfo{person}{Jie Liang}, \bibinfo{person}{Zhiyong Wu}, \bibinfo{person}{Jingzhou Fu}, \bibinfo{person}{Yiyuan Bai}, \bibinfo{person}{Qiang Zhang}, {and} \bibinfo{person}{Yu Jiang}.} \bibinfo{year}{2024}\natexlab{a}.
\newblock \showarticletitle{{WingFuzz}: Implementing Continuous Fuzzing for {DBMSs}}. In \bibinfo{booktitle}{\emph{2024 USENIX Annual Technical Conference (USENIX ATC 24)}}. \bibinfo{publisher}{USENIX Association}, \bibinfo{address}{Santa Clara, CA}, \bibinfo{pages}{479--492}.
\newblock
\showISBNx{978-1-939133-41-0}
\urldef\tempurl%
\url{https://www.usenix.org/conference/atc24/presentation/liang}
\showURL{%
\tempurl}


\bibitem[Liang et~al\mbox{.}(2024b)]%
        {mozi}
\bibfield{author}{\bibinfo{person}{Jie Liang}, \bibinfo{person}{Zhiyong Wu}, \bibinfo{person}{Jingzhou Fu}, \bibinfo{person}{Mingzhe Wang}, \bibinfo{person}{Chengnian Sun}, {and} \bibinfo{person}{Yu Jiang}.} \bibinfo{year}{2024}\natexlab{b}.
\newblock \showarticletitle{Mozi: Discovering DBMS Bugs via Configuration-Based Equivalent Transformation}. In \bibinfo{booktitle}{\emph{Proceedings of the IEEE/ACM 46th International Conference on Software Engineering}} (Lisbon, Portugal) \emph{(\bibinfo{series}{ICSE '24})}. \bibinfo{publisher}{Association for Computing Machinery}, \bibinfo{address}{New York, NY, USA}, Article \bibinfo{articleno}{135}, \bibinfo{numpages}{12}~pages.
\newblock
\showISBNx{9798400702174}
\href{https://doi.org/10.1145/3597503.3639112}{doi:\nolinkurl{10.1145/3597503.3639112}}


\bibitem[Liang et~al\mbox{.}(2022)]%
        {liang2022detecting}
\bibfield{author}{\bibinfo{person}{Yu Liang}, \bibinfo{person}{Song Liu}, {and} \bibinfo{person}{Hong Hu}.} \bibinfo{year}{2022}\natexlab{}.
\newblock \showarticletitle{Detecting Logical Bugs of {DBMS} with Coverage-based Guidance}. In \bibinfo{booktitle}{\emph{31st USENIX Security Symposium (USENIX Security 22)}}. \bibinfo{publisher}{USENIX Association}, \bibinfo{address}{Boston, MA}, \bibinfo{pages}{4309--4326}.
\newblock
\showISBNx{978-1-939133-31-1}
\urldef\tempurl%
\url{https://www.usenix.org/conference/usenixsecurity22/presentation/liang}
\showURL{%
\tempurl}


\bibitem[Liu et~al\mbox{.}(2024)]%
        {liu2024conformance}
\bibfield{author}{\bibinfo{person}{Shuang Liu}, \bibinfo{person}{Chenglin Tian}, \bibinfo{person}{Jun Sun}, \bibinfo{person}{Ruifeng Wang}, \bibinfo{person}{Wei Lu}, \bibinfo{person}{Yongxin Zhao}, \bibinfo{person}{Yinxing Xue}, \bibinfo{person}{Junjie Wang}, {and} \bibinfo{person}{Xiaoyong Du}.} \bibinfo{year}{2024}\natexlab{}.
\newblock \showarticletitle{Conformance Testing of Relational DBMS Against SQL Specifications}.
\newblock \bibinfo{journal}{\emph{arXiv preprint arXiv:2406.09469}} (\bibinfo{year}{2024}).
\newblock


\bibitem[{LLVM Project}(2017)]%
        {libfuzzer}
\bibfield{author}{\bibinfo{person}{{LLVM Project}}.} \bibinfo{year}{2017}\natexlab{}.
\newblock \bibinfo{title}{libFuzzer – a library for coverage-guided fuzz testing}.
\newblock \bibinfo{howpublished}{\url{https://llvm.org/docs/LibFuzzer.html}}.
\newblock
\newblock
\shownote{Accessed: 2024-03-30}.


\bibitem[Marcus et~al\mbox{.}(2019)]%
        {marcus2019neo}
\bibfield{author}{\bibinfo{person}{Ryan Marcus}, \bibinfo{person}{Parimarjan Negi}, \bibinfo{person}{Hongzi Mao}, \bibinfo{person}{Chi Zhang}, \bibinfo{person}{Mohammad Alizadeh}, \bibinfo{person}{Tim Kraska}, \bibinfo{person}{Olga Papaemmanouil}, {and} \bibinfo{person}{Nesime Tatbul}.} \bibinfo{year}{2019}\natexlab{}.
\newblock \showarticletitle{Neo: a learned query optimizer}.
\newblock \bibinfo{journal}{\emph{Proc. VLDB Endow.}} \bibinfo{volume}{12}, \bibinfo{number}{11} (\bibinfo{date}{jul} \bibinfo{year}{2019}), \bibinfo{pages}{1705–1718}.
\newblock
\showISSN{2150-8097}
\href{https://doi.org/10.14778/3342263.3342644}{doi:\nolinkurl{10.14778/3342263.3342644}}


\bibitem[Padhye et~al\mbox{.}(2019)]%
        {padhye2019fuzzing}
\bibfield{author}{\bibinfo{person}{Rohan Padhye}, \bibinfo{person}{Caroline Lemieux}, \bibinfo{person}{Koushik Sen}, \bibinfo{person}{Mike Papadakis}, {and} \bibinfo{person}{Yves Le~Traon}.} \bibinfo{year}{2019}\natexlab{}.
\newblock \showarticletitle{Semantic Fuzzing with Zest}. In \bibinfo{booktitle}{\emph{Proceedings of the 28th ACM SIGSOFT International Symposium on Software Testing and Analysis}} (Beijing, China) \emph{(\bibinfo{series}{ISSTA 2019})}. \bibinfo{publisher}{Association for Computing Machinery}, \bibinfo{address}{New York, NY, USA}, \bibinfo{pages}{329–340}.
\newblock
\showISBNx{9781450362245}
\href{https://doi.org/10.1145/3293882.3330576}{doi:\nolinkurl{10.1145/3293882.3330576}}


\bibitem[PingCAP-Qe({[n.\,d.]})]%
        {PingCAP-Qe}
\bibfield{author}{\bibinfo{person}{PingCAP-Qe}.} \bibinfo{year}{[n.\,d.]}\natexlab{}.
\newblock \bibinfo{title}{GitHub - PingCAP-QE/go-sqlancer: go-sqlancer}.
\newblock
\urldef\tempurl%
\url{https://github.com/PingCAP-QE/go-sqlancer}
\showURL{%
\tempurl}


\bibitem[Rigger and Su(2020a)]%
        {rigger2020detecting}
\bibfield{author}{\bibinfo{person}{Manuel Rigger} {and} \bibinfo{person}{Zhendong Su}.} \bibinfo{year}{2020}\natexlab{a}.
\newblock \showarticletitle{Detecting optimization bugs in database engines via non-optimizing reference engine construction}. In \bibinfo{booktitle}{\emph{Proceedings of the 28th ACM Joint Meeting on European Software Engineering Conference and Symposium on the Foundations of Software Engineering}} (Virtual Event, USA) \emph{(\bibinfo{series}{ESEC/FSE 2020})}. \bibinfo{publisher}{Association for Computing Machinery}, \bibinfo{address}{New York, NY, USA}, \bibinfo{pages}{1140–1152}.
\newblock
\showISBNx{9781450370431}
\href{https://doi.org/10.1145/3368089.3409710}{doi:\nolinkurl{10.1145/3368089.3409710}}


\bibitem[Rigger and Su(2020b)]%
        {rigger2020finding}
\bibfield{author}{\bibinfo{person}{Manuel Rigger} {and} \bibinfo{person}{Zhendong Su}.} \bibinfo{year}{2020}\natexlab{b}.
\newblock \showarticletitle{Finding bugs in database systems via query partitioning}.
\newblock \bibinfo{journal}{\emph{Proc. ACM Program. Lang.}} \bibinfo{volume}{4}, \bibinfo{number}{OOPSLA}, Article \bibinfo{articleno}{211} (\bibinfo{date}{nov} \bibinfo{year}{2020}), \bibinfo{numpages}{30}~pages.
\newblock
\href{https://doi.org/10.1145/3428279}{doi:\nolinkurl{10.1145/3428279}}


\bibitem[Rigger and Su(2020c)]%
        {rigger2020testing}
\bibfield{author}{\bibinfo{person}{Manuel Rigger} {and} \bibinfo{person}{Zhendong Su}.} \bibinfo{year}{2020}\natexlab{c}.
\newblock \showarticletitle{Testing database engines via pivoted query synthesis}. In \bibinfo{booktitle}{\emph{14th USENIX Symposium on Operating Systems Design and Implementation (OSDI 20)}}. \bibinfo{pages}{667--682}.
\newblock


\bibitem[Seltenreich(2022)]%
        {sqlsmith}
\bibfield{author}{\bibinfo{person}{Andreas Seltenreich}.} \bibinfo{year}{2022}\natexlab{}.
\newblock \bibinfo{title}{Sqlsmith}.
\newblock
\urldef\tempurl%
\url{https://github.com/anse1/sqlsmith}
\showURL{%
\tempurl}


\bibitem[Serebryany(2017)]%
        {kostya2017oss}
\bibfield{author}{\bibinfo{person}{Kostya Serebryany}.} \bibinfo{year}{2017}\natexlab{}.
\newblock \showarticletitle{{OSS-Fuzz} - Google{\textquoteright}s continuous fuzzing service for open source software}. \bibinfo{publisher}{USENIX Association}, \bibinfo{address}{Vancouver, BC}.
\newblock


\bibitem[Serebryany et~al\mbox{.}(2012)]%
        {konstantin2012addresssanitizer}
\bibfield{author}{\bibinfo{person}{Konstantin Serebryany}, \bibinfo{person}{Derek Bruening}, \bibinfo{person}{Alexander Potapenko}, {and} \bibinfo{person}{Dmitriy Vyukov}.} \bibinfo{year}{2012}\natexlab{}.
\newblock \showarticletitle{{AddressSanitizer}: A Fast Address Sanity Checker}. In \bibinfo{booktitle}{\emph{2012 USENIX Annual Technical Conference (USENIX ATC 12)}}. \bibinfo{publisher}{USENIX Association}, \bibinfo{address}{Boston, MA}, \bibinfo{pages}{309--318}.
\newblock
\showISBNx{978-931971-93-5}
\urldef\tempurl%
\url{https://www.usenix.org/conference/atc12/technical-sessions/presentation/serebryany}
\showURL{%
\tempurl}


\bibitem[Serebryany and Iskhodzhanov(2009)]%
        {serebryany2009threadsanitizer}
\bibfield{author}{\bibinfo{person}{Konstantin Serebryany} {and} \bibinfo{person}{Timur Iskhodzhanov}.} \bibinfo{year}{2009}\natexlab{}.
\newblock \showarticletitle{ThreadSanitizer: data race detection in practice}. In \bibinfo{booktitle}{\emph{Proceedings of the Workshop on Binary Instrumentation and Applications}} (New York, New York, USA) \emph{(\bibinfo{series}{WBIA '09})}. \bibinfo{publisher}{Association for Computing Machinery}, \bibinfo{address}{New York, NY, USA}, \bibinfo{pages}{62–71}.
\newblock
\showISBNx{9781605587936}
\href{https://doi.org/10.1145/1791194.1791203}{doi:\nolinkurl{10.1145/1791194.1791203}}


\bibitem[Sidler et~al\mbox{.}(2017)]%
        {sidler2017doppiodb}
\bibfield{author}{\bibinfo{person}{David Sidler}, \bibinfo{person}{Zsolt Istvan}, \bibinfo{person}{Muhsen Owaida}, \bibinfo{person}{Kaan Kara}, {and} \bibinfo{person}{Gustavo Alonso}.} \bibinfo{year}{2017}\natexlab{}.
\newblock \showarticletitle{doppioDB: A Hardware Accelerated Database}. In \bibinfo{booktitle}{\emph{Proceedings of the 2017 ACM International Conference on Management of Data}} (Chicago, Illinois, USA) \emph{(\bibinfo{series}{SIGMOD '17})}. \bibinfo{publisher}{Association for Computing Machinery}, \bibinfo{address}{New York, NY, USA}, \bibinfo{pages}{1659–1662}.
\newblock
\showISBNx{9781450341974}
\href{https://doi.org/10.1145/3035918.3058746}{doi:\nolinkurl{10.1145/3035918.3058746}}


\bibitem[Slutz(1998)]%
        {slutz1998massive}
\bibfield{author}{\bibinfo{person}{Donald~R. Slutz}.} \bibinfo{year}{1998}\natexlab{}.
\newblock \showarticletitle{Massive Stochastic Testing of SQL}. In \bibinfo{booktitle}{\emph{Proceedings of the 24rd International Conference on Very Large Data Bases}} \emph{(\bibinfo{series}{VLDB '98})}. \bibinfo{publisher}{Morgan Kaufmann Publishers Inc.}, \bibinfo{address}{San Francisco, CA, USA}, \bibinfo{pages}{618–622}.
\newblock
\showISBNx{1558605665}


\bibitem[Song et~al\mbox{.}(2023)]%
        {song2023testing}
\bibfield{author}{\bibinfo{person}{Jiansen Song}, \bibinfo{person}{Wensheng Dou}, \bibinfo{person}{Ziyu Cui}, \bibinfo{person}{Qianwang Dai}, \bibinfo{person}{Wei Wang}, \bibinfo{person}{Jun Wei}, \bibinfo{person}{Hua Zhong}, {and} \bibinfo{person}{Tao Huang}.} \bibinfo{year}{2023}\natexlab{}.
\newblock \showarticletitle{Testing Database Systems via Differential Query Execution}. In \bibinfo{booktitle}{\emph{2023 IEEE/ACM 45th International Conference on Software Engineering (ICSE)}}. \bibinfo{pages}{2072--2084}.
\newblock
\href{https://doi.org/10.1109/ICSE48619.2023.00175}{doi:\nolinkurl{10.1109/ICSE48619.2023.00175}}


\bibitem[Song et~al\mbox{.}(2024)]%
        {song2024detecting}
\bibfield{author}{\bibinfo{person}{Jiansen Song}, \bibinfo{person}{Wensheng Dou}, \bibinfo{person}{Yu Gao}, \bibinfo{person}{Ziyu Cui}, \bibinfo{person}{Yingying Zheng}, \bibinfo{person}{Dong Wang}, \bibinfo{person}{Wei Wang}, \bibinfo{person}{Jun Wei}, {and} \bibinfo{person}{Tao Huang}.} \bibinfo{year}{2024}\natexlab{}.
\newblock \showarticletitle{Detecting Metadata-Related Logic Bugs in Database Systems via Raw Database Construction}.
\newblock \bibinfo{journal}{\emph{Proc. VLDB Endow.}} \bibinfo{volume}{17}, \bibinfo{number}{8} (\bibinfo{date}{may} \bibinfo{year}{2024}), \bibinfo{pages}{1884–1897}.
\newblock
\showISSN{2150-8097}
\href{https://doi.org/10.14778/3659437.3659445}{doi:\nolinkurl{10.14778/3659437.3659445}}


\bibitem[Stepanov and Serebryany(2015)]%
        {stepanov2015memorysanitizer}
\bibfield{author}{\bibinfo{person}{Evgeniy Stepanov} {and} \bibinfo{person}{Konstantin Serebryany}.} \bibinfo{year}{2015}\natexlab{}.
\newblock \showarticletitle{MemorySanitizer: Fast detector of uninitialized memory use in C++}. In \bibinfo{booktitle}{\emph{2015 IEEE/ACM International Symposium on Code Generation and Optimization (CGO)}}. \bibinfo{pages}{46--55}.
\newblock
\href{https://doi.org/10.1109/CGO.2015.7054186}{doi:\nolinkurl{10.1109/CGO.2015.7054186}}


\bibitem[Tang et~al\mbox{.}(2023)]%
        {tang2023detecting}
\bibfield{author}{\bibinfo{person}{Xiu Tang}, \bibinfo{person}{Sai Wu}, \bibinfo{person}{Dongxiang Zhang}, \bibinfo{person}{Feifei Li}, {and} \bibinfo{person}{Gang Chen}.} \bibinfo{year}{2023}\natexlab{}.
\newblock \showarticletitle{Detecting Logic Bugs of Join Optimizations in DBMS}.
\newblock \bibinfo{journal}{\emph{Proc. ACM Manag. Data}} \bibinfo{volume}{1}, \bibinfo{number}{1}, Article \bibinfo{articleno}{55} (\bibinfo{date}{may} \bibinfo{year}{2023}), \bibinfo{numpages}{26}~pages.
\newblock
\href{https://doi.org/10.1145/3588909}{doi:\nolinkurl{10.1145/3588909}}


\bibitem[Vitess(2024)]%
        {vitess2024fuzzing}
\bibfield{author}{\bibinfo{person}{Vitess}.} \bibinfo{year}{2024}\natexlab{}.
\newblock \bibinfo{title}{Vitess Fuzzing Summer 2023 Internship}.
\newblock \bibinfo{howpublished}{\url{https://vitess.io/blog/2024-04-08-vitess-fuzzing-summer-2023-internship/}}.
\newblock
\newblock
\shownote{Accessed: 2024-10-09}.


\bibitem[Yang et~al\mbox{.}(2024)]%
        {yang2024towards}
\bibfield{author}{\bibinfo{person}{Yupeng Yang}, \bibinfo{person}{Yongheng Chen}, \bibinfo{person}{Rui Zhong}, \bibinfo{person}{Jizhou Chen}, {and} \bibinfo{person}{Wenke Lee}.} \bibinfo{year}{2024}\natexlab{}.
\newblock \showarticletitle{Towards Generic Database Management System Fuzzing}. In \bibinfo{booktitle}{\emph{33rd USENIX Security Symposium (USENIX Security 24)}}. \bibinfo{publisher}{USENIX Association}, \bibinfo{address}{Philadelphia, PA}, \bibinfo{pages}{901--918}.
\newblock
\showISBNx{978-1-939133-44-1}
\urldef\tempurl%
\url{https://www.usenix.org/conference/usenixsecurity24/presentation/yang-yupeng}
\showURL{%
\tempurl}


\bibitem[Zalewski(2020)]%
        {AFL}
\bibfield{author}{\bibinfo{person}{Michal Zalewski}.} \bibinfo{year}{2020}\natexlab{}.
\newblock \bibinfo{title}{American Fuzzy Lop (AFL)}.
\newblock \bibinfo{howpublished}{https://github.com/google/AFL}.
\newblock
\newblock
\shownote{Accessed: 2024-04-16}.


\bibitem[Zhang et~al\mbox{.}(2023)]%
        {zhang2023understanding}
\bibfield{author}{\bibinfo{person}{Cen Zhang}, \bibinfo{person}{Mingqiang Bai}, \bibinfo{person}{Yaowen Zheng}, \bibinfo{person}{Yeting Li}, \bibinfo{person}{Xiaofei Xie}, \bibinfo{person}{Yuekang Li}, \bibinfo{person}{Wei Ma}, \bibinfo{person}{Limin Sun}, {and} \bibinfo{person}{Yang Liu}.} \bibinfo{year}{2023}\natexlab{}.
\newblock \bibinfo{title}{Understanding Large Language Model Based Fuzz Driver Generation}.
\newblock
\showeprint[arxiv]{2307.12469}~[cs.CR]


\bibitem[Zhang and Rigger(2025)]%
        {zhang2025constant}
\bibfield{author}{\bibinfo{person}{Chi Zhang} {and} \bibinfo{person}{Manuel Rigger}.} \bibinfo{year}{2025}\natexlab{}.
\newblock \showarticletitle{Constant Optimization Driven Database System Testing}.
\newblock \bibinfo{journal}{\emph{Proc. ACM Manag. Data}} \bibinfo{volume}{3}, \bibinfo{number}{1}, Article \bibinfo{articleno}{24} (\bibinfo{date}{Feb.} \bibinfo{year}{2025}), \bibinfo{numpages}{24}~pages.
\newblock
\href{https://doi.org/10.1145/3709674}{doi:\nolinkurl{10.1145/3709674}}


\bibitem[Zhang et~al\mbox{.}(2017)]%
        {zhang2017skeletal}
\bibfield{author}{\bibinfo{person}{Qirun Zhang}, \bibinfo{person}{Chengnian Sun}, {and} \bibinfo{person}{Zhendong Su}.} \bibinfo{year}{2017}\natexlab{}.
\newblock \showarticletitle{Skeletal Program Enumeration for Rigorous Compiler Testing}.
\newblock \bibinfo{journal}{\emph{SIGPLAN Not.}} \bibinfo{volume}{52}, \bibinfo{number}{6} (\bibinfo{date}{jun} \bibinfo{year}{2017}), \bibinfo{pages}{347–361}.
\newblock
\showISSN{0362-1340}
\href{https://doi.org/10.1145/3140587.3062379}{doi:\nolinkurl{10.1145/3140587.3062379}}


\bibitem[Zhong et~al\mbox{.}(2020)]%
        {zhong2020squirrel}
\bibfield{author}{\bibinfo{person}{Rui Zhong}, \bibinfo{person}{Yongheng Chen}, \bibinfo{person}{Hong Hu}, \bibinfo{person}{Hangfan Zhang}, \bibinfo{person}{Wenke Lee}, {and} \bibinfo{person}{Dinghao Wu}.} \bibinfo{year}{2020}\natexlab{}.
\newblock \showarticletitle{SQUIRREL: Testing Database Management Systems with Language Validity and Coverage Feedback}. In \bibinfo{booktitle}{\emph{Proceedings of the 2020 ACM SIGSAC Conference on Computer and Communications Security}} (Virtual Event, USA) \emph{(\bibinfo{series}{CCS '20})}. \bibinfo{publisher}{Association for Computing Machinery}, \bibinfo{address}{New York, NY, USA}, \bibinfo{pages}{955–970}.
\newblock
\showISBNx{9781450370899}
\href{https://doi.org/10.1145/3372297.3417260}{doi:\nolinkurl{10.1145/3372297.3417260}}


\end{thebibliography}

\appendix

\section{Implementation}
We illustrate the key implementation details of the adaptive statement generator.
\subsection{Supported Features}\label{sec:implementation}
We summarize the SQL features supported by our generator.
We considered basic and mostly standardized SQL features that we believe could successfully execute on different DBMSs.
We summarize and explain the features in the following paragraphs; the exact set of features is detailed in the artifact. 
We have identified four common concrete granularities of \emph{SQL features}, namely
\emph{statements}, \emph{clauses}, expressions (\emph{functions} and \emph{operators}), \emph{data types} as well as \emph{abstract properties} (see Table~\ref{tab:sql-features}).

We overall implemented only six common statements in the generator, including \texttt{CREATE TABLE}, \texttt{CREATE INDEX}, \texttt{CREATE VIEW}, \texttt{INSERT}, \texttt{ANALYZE}, and \texttt{SELECT}.
Note that even such basic statements are not supported by all DBMSs (\emph{e.g.}, CrateDB does not support \texttt{CREATE INDEX}).

Statements often have some optional \emph{keywords} or \emph{clauses} associated with them, such as \texttt{UNIQUE} and \texttt{INNER JOIN} in the above example. 
The keywords correspond to concrete strings in the statements, and the clauses usually contain various expressions.
We assume the \texttt{WHERE} clause should be supported by every DBMS. 
However, some types of join clauses are not, for example, \texttt{RIGHT JOIN} was only supported in SQLite since 2022.
We support six types of join.

\begin{figure}[tb]
    \centering
    \includegraphics[width=\linewidth]{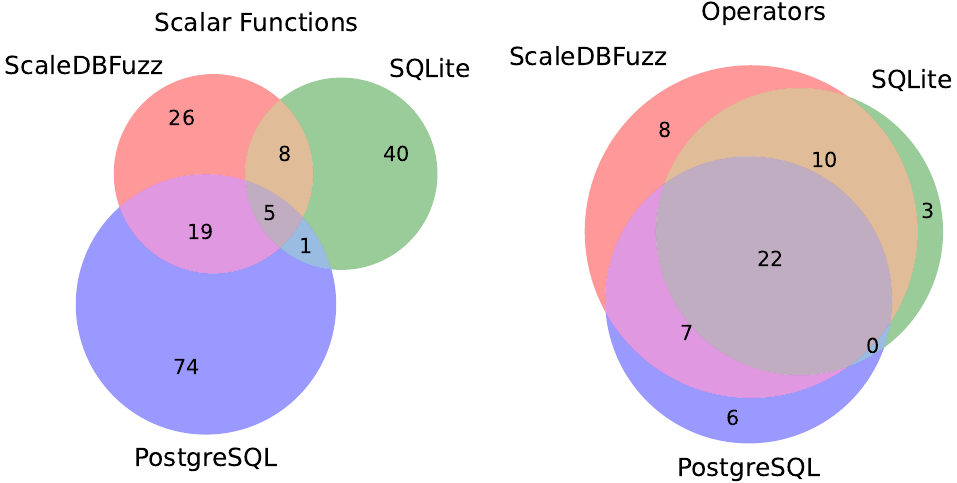}
    \caption{Venn diagram of SQL features shared by \ProjectName{}, and the SQLite and PostgreSQL \SQLancer{} generators.}
    \label{fig:venn}
\end{figure}

Expressions consist of various \emph{operators}, \emph{functions}, as well as constants and column references, such as the comparison operators (\texttt{=}) and math function (\texttt{SIN}) shown in the table.
Expressions are used in various contexts, such as in \texttt{WHERE} clauses or in \texttt{ON} for \texttt{JOIN} clauses.
In total, we support \FunctionNodes{} functions and \OperatorNodes{} operators.
Figure~\ref{fig:venn} shows the number of distinct and shared features implemented in \ProjectName{} and the SQLite and PostgreSQL SQLancer generators.
The leaf nodes of the expressions are constants and column references, which can have different \emph{data type}s.
\ProjectName{} supports generating three data types: integer, string, and boolean.

With respect to \emph{abstract properties}, they mostly relate to the type system supported by the DBMS.
Most importantly, we consider whether the DBMS' SQL dialect is statically typed or dynamically typed, and we represent these as features.
For example, PostgreSQL is statically typed as it provides few implicit casts and rejects ill-typed statements, while SQLite is dynamically typed, as it can coerce most values to the required data type at run time.
This influences, for example, the generation of expressions.
Consider a \texttt{WHERE} clause in a \texttt{SELECT} statement.
For an untyped DBMS, the expression generator is free to generate an expression of any type.
For a strictly typed system, it can only choose operators that produce a boolean expression; if this expression is, for example, a comparison operator, it must ensure that the compared operands have a compatible type.
Rather than hard-coding the operand types of operators and argument types in functions, we also provide finer-grained features that learn the expected types.
See Figure~\ref{fig:feedback}, where the identifier \texttt{SIN1INT} represents that the first argument of the function \texttt{SIN} is type integer.
Further, if the generator generates \texttt{SIN(`a')}, it will be captured as a feature \texttt{SIN1STRING}.
Note that further properties are plausible; for example, we noticed that some features cannot co-occur, or some features require other features to be present. We currently do not capture such complex relationships.

\begin{table}[tb]
    \centering
\small
\caption{SQL features}
\label{tab:sql-features}
\vspace{-3mm}
\begin{tabular}{llrl}
\toprule
\multicolumn{2}{c}{Feature Type}  & Number & Examples \\

\midrule
\multicolumn{2}{l}{Statement} & 6    &   \texttt{CREATE INDEX}   \\
\multicolumn{2}{l}{Clause \& Keyword}     & 10   & \texttt{RIGHT JOIN}, \texttt{SUBQUERY} \\
\multirow{2}{*}{Expression}     &  Function   & \FunctionNodes{}  &  \texttt{NULLIF}, \texttt{SIN}    \\
                                 &  Operator   & \OperatorNodes{}  &  \texttt{+}, \texttt{=}, \texttt{AND}, \texttt{CASE-WHEN}   \\
\multicolumn{2}{l}{Data type}   & 3    &      \texttt{INTEGER}      \\
\bottomrule
\end{tabular}

\end{table}
\begin{figure}[tb]

\captionof{lstlisting}{Illustrative example of the adaptive statement generator. Design differences with non-adaptive generators are highlighted in red (non-adaptive) and green (adaptive).}
\label{listing:general-expr}

\begin{minipage}{.45\textwidth}
    \centering
    \lstset{language=Java,xleftmargin=2em, lineskip=0pt, aboveskip=0pt,belowskip=0pt, basicstyle=\ttfamily\footnotesize}
\begin{lstlisting}[firstline=1, lastline=3]
public SQLStatement createIndex(GeneralState state) {
  //...
  String stmt = "CREATE ";
\end{lstlisting}
\begin{lstlisting}[firstline=4, lastline=4,firstnumber=4,backgroundcolor=\color{red!30} ]
-  if (state.nextBoolean()) {
\end{lstlisting}
\begin{lstlisting}[firstline=5, lastline=6,firstnumber=5,backgroundcolor=\color{green!30}  ]
+  if (state.shouldGenerate(Node.INDEX_UNIQUE) 
+   && state.nextBoolean()) {
\end{lstlisting}
\begin{lstlisting}[firstline=7, lastline=7,firstnumber=7, ]
    stmt += "UNIQUE ";
\end{lstlisting}
\begin{lstlisting}[firstline=8, firstnumber=8, lastline=8,backgroundcolor=\color{green!30}]
+    state.generateFeature(Node.INDEX_UNIQUE);
\end{lstlisting}
\begin{lstlisting}[firstline=9, firstnumber=9]
  }
  //...
  return new SQLStatement(stmt, errors);
}
\end{lstlisting}
\end{minipage}

\end{figure}

\subsection{Feedback mechanism}
Listing~\ref{listing:general-expr} demonstrates the simplified index generator of the adaptive generator, highlighting implementation differences compared to the non-adaptive one in \SQLancer{}. We implemented the other generators similarly.
When the generator selects new random alternatives, including statement keywords (line 7) and expression nodes, the feedback mechanisms indicate whether the option is supported, and record the feature selection for subsequent feedback updates.
In addition to the interface shown in the example above, alternative interfaces allow querying the feedback generator for one of multiple supported features.
In line 5, if the initial execution threshold has been reached, and the feature was found to not be supported by the DBMS under test, the call to \texttt{shouldGenerate} returns false.
In line 8, we instruct the generator that we have selected the feature for generation. After execution of the SQL statement returned by line 11, the generator will use this information to update the validity rate.
From the perspective of programmers extending \ProjectName{}'s generator, the internals are thus mostly transparent.

\subsection{Execution strategies}
As with many other automated testing approaches, we have empirically determined suitable thresholds and strategies for our generator.
First, the generator begins by generating expressions with low depth and gradually increases it.
Specifically, the generator starts generating expressions at a depth of 1, and increases the depth after each $I$ executions by 1, up to a depth of 3.
A lower depth corresponds to a reduced number of features, which enhances the learning efficiency of the generator since multiple features in one statement make it difficult to isolate the unsupported features.
Besides, this enables it to identify simple bug-inducing features early and potentially deprioritize more complex ones later.
Second, the iterations $I$ of executing test cases should be sufficient. The generator updates the probabilistic rules for each 100K test case, and we believe the number enables each feature to be sufficiently often executed. A larger $I$ should be set if more features are incorporated or the probability threshold is more strict (\emph{e.g.}, less than 1$\%$).

\section{Artifact Appendix}

\subsection{Abstract}

This artifact contains the SQLancer++ implementation, Dockerized DBMS setups, and reproduction scripts.
It supports reproducing the main results reported in the paper, including:
bug-triggering test cases for each reported bug (Table~\ref{table:DBMSs}),
the SQL feature study (Figure~\ref{fig:featureheatmap}),
coverage and feedback effectiveness (Tables~\ref{tab:coverage}--\ref{tab:validity}),
and the bug prioritization evaluation (Table~\ref{tab:deduplication}).
Most experiments are executed inside Docker containers for isolation, although the experiment for Section~\ref{sec:featurestudy} runs directly on the host machine.

\subsection{Artifact check-list (meta-information)}

{\small
\begin{itemize}
  \item {\bf Program: } \texttt{SQLancerPlusPlus/} (Java; Maven build)
  \item {\bf Compilation: } Maven (\texttt{mvn clean package -DskipTests})
  \item {\bf Run-time environment: } Linux; Docker
  \item {\bf Hardware: } x86\_64; at least 32 cores, 64\,GB RAM
  \item {\bf Metrics: } Unique bugs found, code coverage, and validity rate.
  \item {\bf Output: } \texttt{logs/} and generated plots/tables
  \item {\bf Experiments: } Reproduction for Sections~\ref{sec:featurestudy}--\ref{sec: bug-deduplication}
  \item {\bf How much disk space required (approximately)?: } 50 GB
  \item {\bf How much time is needed to prepare workflow (approximately)?: } About 1--2 hours
  \item {\bf How much time is needed to complete experiments (approximately)?: } Around 1 day with parallel execution, or 2--3 days sequentially
  \item {\bf Publicly available?: } Yes
  \item {\bf Code licenses (if publicly available)?: } MIT
  \item {\bf Archived (provide DOI)?: } \url{https://doi.org/10.5281/zenodo.18289297}
\end{itemize}
}

\subsection{Description}

\subsubsection{How to access}

All materials can be downloaded from \url{https://doi.org/10.5281/zenodo.18289297}.

\subsubsection{Hardware dependencies}

An x86 CPU with at least 64 GB of RAM is required.
Having more cores speeds up the experiments through increased parallelism.
We used a 64-core AMD EPYC 7763 CPU at 2.45GHz and 512GB memory for our experiments.

\subsubsection{Software Dependencies}

Our artifact requires Java~11 (or later), Python~3.10 (or later), Docker, and Maven.

\subsubsection{Data sets}

The bug list is included in the artifact under the \texttt{bugs} directory.

\subsection{Installation}

The installation steps and scripts assume a recent version of Ubuntu (\emph{e.g.}, 22.04) and that the software dependencies (Java, Maven, Docker, and Python) above are already installed.

\begin{enumerate}
    \item \textbf{Build SQLancer++:} 
\begin{verbatim}
cd SQLancerPlusPlus
mvn clean package -DskipTests
\end{verbatim}
    \item \textbf{Build Docker images:}
\begin{verbatim}
./prepare_experiments_images.sh
\end{verbatim}
    \item \textbf{Prepare Python environment:}
\begin{verbatim}
pip install matplotlib numpy pandas
\end{verbatim}
\end{enumerate}

\subsection{Experiment workflow}

The experiments are launched by scripts on the host OS.
These scripts start multiple Docker containers in isolation, save all collected results under the \texttt{logs} directory, and generate the tables and figures used in the paper.
A typical workflow is:
\begin{enumerate}
    \item Read \texttt{README.md} for an overview.
    \item Build the Docker images.
    \item Inspect the bug reports (for Section~\ref{sec: bug-detect}).
    \item Analyze the bug-inducing test cases (for Section~\ref{sec:featurestudy}).
    \item Run the experiments (for Sections~\ref{sec: bug-efficiency}--\ref{sec: bug-deduplication}) and inspect the results.
\end{enumerate}

\subsection{Evaluation and expected results}

The expected outputs include:
\begin{itemize}
  \item \textbf{SQL feature study:} \texttt{bugs/feature\_heatmap.pdf}
  \item \textbf{Coverage and validity rate:} the execution logs under \texttt{logs/}, and the generated files \texttt{coverage.tex} (for Table~\ref{tab:coverage}) and \texttt{validity.tex} (for Table~\ref{tab:validity})
  \item \textbf{Bug prioritization:} CrateDB logs and result files under \texttt{logs/}, used for Table~\ref{tab:deduplication}
\end{itemize}

\subsection{Experiment customization}

The provided scripts assume parallel execution with fixed time budgets as in the original paper, but they accept parameters that allow adjusting time budgets and switching to sequential execution.

\subsection{Methodology}

Submission, reviewing, and badging methodology:
\begin{itemize}
  \item \url{https://www.acm.org/publications/policies/artifact-review-and-badging-current}
  \item \url{https://cTuning.org/ae}
\end{itemize}

\end{document}